\documentclass[12pt,preprint]{aastex}
\newcommand{\asec}      {\mbox{$^{\prime \prime}  $} }

\newcommand{\etal}{{\it et al.}}

\newcommand{\bq}{\begin{equation}}
\newcommand{\eq}{\end{equation}}

\shorttitle{Photometric Redshifts of Galaxies in COSMOS}
\shortauthors{Mobasher {\it B. Mobasher et al}}

\begin{document}

\title{Photometric Redshifts of Galaxies in COSMOS$^1$}
\author{
Mobasher, B.\altaffilmark{2};
Capak, P. \altaffilmark{3};
Scoville, N. Z. \altaffilmark{3};
Dahlen, T \altaffilmark{4};
Salvato, M. \altaffilmark{3};
Aussel, H.\altaffilmark{5};
Thompson, D. J. \altaffilmark{6}
Feldmann, R. \altaffilmark{7};
Tasca, L. \altaffilmark{8};
Lefevre, O. \altaffilmark{8};
Lilly, S. \altaffilmark{7};
Carollo, C. M. \altaffilmark{7};
Kartaltepe, J. S. \altaffilmark{12};
McCracken, H. \altaffilmark{9};
Mould, J. \altaffilmark{10};
Renzini, A. \altaffilmark{11};
Sanders, D. B. \altaffilmark{12};
Shopbell, P. L. \altaffilmark{3};
Taniguchi, Y. \altaffilmark{13};
Ajiki, M. \altaffilmark{13};
Shioya, Y. \altaffilmark{13};
Contini, T. \altaffilmark{14};
Giavalisco,M. \altaffilmark{2};
Ilbert, O. \altaffilmark{12};
Iovino, A. \altaffilmark{15};
Le Brun, V. \altaffilmark{8};
Mainieri, V. \altaffilmark{16};
Mignoli, M. \altaffilmark{17};
Scodeggio, M. \altaffilmark{18}
}
\altaffiltext{$\star$}{Based on observations with the NASA/ESA {\em
Hubble Space Telescope}, obtained at the Space Telescope Science
Institute, which is operated by AURA Inc, under NASA contract NAS
5-26555; also based on data collected at : the Subaru Telescope, which is operat
ed by
the National Astronomical Observatory of Japan; the XMM-Newton, an ESA science m
ission with
instruments and contributions directly funded by ESA Member States and NASA; the
 European Southern Observatory under Large Program 175.A-0839, Chile; Kitt Peak 
National Observatory, Cerro Tololo Inter-American
Observatory, and the National Optical Astronomy Observatory, which are
operated by the Association of Universities for Research in Astronomy, Inc.
(AURA) under cooperative agreement with the National Science Foundation; 
the National Radio Astronomy Observatory which is a facility of the National Sci
ence 
Foundation operated under cooperative agreement by Associated Universities, Inc 
; 
and and the Canada-France-Hawaii Telescope with MegaPrime/MegaCam operated as a
joint project by the CFHT Corporation, CEA/DAPNIA, the National Research
Council of Canada, the Canadian Astronomy Data Centre, the Centre National
de la Recherche Scientifique de France, TERAPIX and the University of
Hawaii.}  
\altaffiltext{2}{Space Telescope Science Institute, 3700 San Martin
Drive, Baltimore, MD 21218}
\altaffiltext{3}{California Institute of Technology, MC 105-24, 1200 East
California Boulevard, Pasadena, CA 91125}
\altaffiltext{4}{Department of Physics, Stockholm University, SE-10961, Stockholm, Sweden}
\altaffiltext{5}{Service d'Astrophysique, CEA/Saclay, 91191 Gif-sur-Yvette, France}
\altaffiltext{6}{Large Binocular Telescope Observatory, University of Arizona
, 933 N. Cherry Ave.  Tucson, AZ  85721-0065,   USA}
\altaffiltext{7}{Department of Physics, ETH Zurich, CH-8093 Zurich, Switzerlan
d}
\altaffiltext{8}{Laboratoire d'Astrophysique de Marseille, BP 8, Traverse
du Siphon, 13376 Marseille Cedex 12, France}
\altaffiltext{9}{Institut d'Astrophysique de Paris, UMR7095 CNRS, Universit\`
e Pierre et Marie Curie, 98 bis Boulevard Arago, 75014 Paris, France}
\altaffiltext{10}{National Optical Astronomy Observatory, P.O. Box 26732, Tucson, AZ 85726}
\altaffiltext{11}{Dipartimento di Astronomia, Universit di Padova, vicolo dell'Osservatorio 2, I-35122 Padua, Italy}
\altaffiltext{12}{Institute for Astronomy, 2680 Woodlawn Dr., University of Hawaii, Honolulu, Hawaii, 96822}
\altaffiltext{13}{Physics Department, Graduate School of Science, Ehime University, 2-5 Bunkyou, Matuyama, 790-8577, Japan}
\altaffiltext{14}{Observatoire Midi-Pyranes, 14 avenue E. Belin, 31400 Toulouse, France}
\altaffiltext{15}{INAF, Osservatorio Astronomico di Brera, via Brera 28, 20121 Milano, Italy}
\altaffiltext{16}{Max Planck Institut fur Extraterrestrische Physik, Garching, Germany}
\altaffiltext{17}{INAF, Osservatorio Astronomico di Bologna, via Ranzani 1, 40127 Bologna, Italy}
\altaffiltext{18}{INAF - IASF Milano, via Bassini 15, 20133 Milano, Italy}

\begin{abstract}

We present photometric redshifts for the COSMOS survey derived from 
a new code, optimized to yield accurate and reliable redshifts and 
spectral types of galaxies down to faint magnitudes and redshifts 
out to $z \sim 1.2$. The technique uses $\chi^2$ template fitting, 
combined with luminosity function priors
and with the option  to estimate the internal extinction (or $E(B-V)$). 
The median most-probable redshift, best-fit spectral type and reddening, 
absolute magnitude and stellar mass are derived in addition to the full 
redshift probability distributions. Using simulations with sampling and noise
similar to those in COSMOS, the accuracy and reliability is 
estimated for the photometric redshifts as a function of the  
magnitude limits of the sample, S/N ratios and the number of bands used.
We find from the simulations that the ratio of derived 95\% confidence 
interval in the $\chi^2$ probability distribution to the estimated 
photometric redshift ($D_{95}$) can be used to identify and exclude the 
catastrophic failures in the photometric redshift estimates. 

To evaluate the reliability of the photometric redshifts, we compare 
the derived redshifts with high-reliability spectroscopic redshifts for 
a sample of 868 normal galaxies with $z < 1.2$ from $z$COSMOS. Considering 
different scenarios, depending on using prior, no prior and/or extinction, 
we compare the photometric and spectroscopic redshifts for this sample. 
The {\it rms} scatter between the estimated photometric redshifts and
known spectroscopic redshifts is $\sigma(\Delta(z))=0.031$, 
where $\Delta(z)=(z_{phot} - z_{spec})/(1 + z_{spec})$ with a small
fraction of outliers ($<2.5\%$)- (outliers are defined as objects with  
$\Delta(z) > 3 \sigma (\Delta(z))$ where $\sigma (\Delta(z))$ is the {\it rms}
scatter in $\Delta(z)$).  We also find good agreement
($\sigma (\Delta(z))=0.10$) between photometric and spectroscopic redshifts 
for Type II AGNs.

We compare results from our photometric redshift procedure with three other 
independent codes and find them in excellent agreement. We show
preliminary results, based on photometric redshifts for the entire COSMOS 
sample (to $i < 25$ mag.).

\end{abstract}


\section{INTRODUCTION}

The determination of galaxy redshifts is a prerequisite to 
studies of their cosmological evolution- measuring both distance-dependent 
quantities such as luminosities, masses and star formation rates and in 
specifying the lookback times. Redshifts are also necessary to separate
out large scale structures and galaxies along the line of sight. 
 With the advent of new 
sensitive detectors on large ground-based telescopes (Subaru, VLT, Keck) 
and space-borne facilities (HST, Spitzer, GALEX), we have now been able to
perform extensive galaxy surveys to unprecedented depths.  
Measurement of spectroscopic redshifts to these galaxies is limited by
two factors: their faintness (Papovich \etal\   2006; Mobasher \etal\   2005;
Yan \etal\   2005) and the large number of galaxies for which such information 
is needed (Wolf \etal\   2004; Mobasher \etal\   2004; Ilbert \etal\  . 2006; 
Salvato \etal\   2006). 

Recently, photometric redshifts have been used extensively 
in deep cosmological surveys, yielding the galaxy luminosity functions 
(Dahlen \etal\   2005; Caputi \etal\   2005) and the 
evolution of star formation rates (Gabasch \etal\   2004; Giavalisco \etal\   2004; Dahlen \etal\   2006). 
The photometric redshift technique has the advantage of providing redshifts
for large samples of faint galaxies with a relatively modest investment
in observing time. For maximal success with photometric redshifts the
photometry should cover as wide a range in wavelength as possible. 
The principle disadvantage of the photometric redshifts
is the relatively low 
resolution in wavelength and redshift (due to the width of filters) 
compared to spectroscopic redshifts. Photometric redshifts are, however, 
 vital in resolving redshift ambiguities where spectroscopy 
shows only a single spectral line (Lilly \etal\   2006). 

In this paper we present measurements of photometric 
redshifts for galaxies in the Cosmic Evolution Survey (COSMOS) and
explore the accuracy of the photometric redshifts based on 
extensive simulations, comparison
with spectroscopic redshifts from zCOSMOS (Lilly \etal\   2006) and 
with photometric redshifts estimated from a number of other 
independent algorithms. Over the $1.4^\circ \times 1.4^\circ$ area
covered by COSMOS, we detect 
367,000 galaxies to $i\sim 25$ (Capak \etal\   2006), making it
difficult to obtain spectroscopic redshifts for the entire galaxy sample.
Extensive multi-waveband photometric data are now available for these 
galaxies, allowing measurement of photometric redshifts for a complete
sample. These results are used to 
identify the large scale structures (Scoville \etal\   2006; 
Finoguenov \etal\   2006; Guzzo \etal\   2006), to study the evolution
of density-morphology relation (Capak \etal\   2006), dependence of the 
star formation activity on the environment (Mobasher \etal\   2006) and 
study of morphologies and rest-frame properties of individual galaxies
(Scarlatta \etal\   2006; Zamojski \etal\   2006).

We present the photometric redshift technique in \S2 followed by
the photometric observations and photometric data in \S3. In \S4 we present 
simulations to explore the dependence of photometric redshifts to 
the magnitude limit, photometric accuracy and S/N ratios. We compare 
photometric and spectroscopic redshifts to a sample of galaxies with 
available such data in \S5. In \S6 we compare results from various
photometric redshift codes. We summarise the galaxy properties derived from 
the photometric data, including SED types and stellar mass measurements
in \S7. 

In this paper we use the standard cosmology with $\Omega_M=0.3$, 
$\Omega_\Lambda=0.7$, and $h=0.7$. Magnitudes are given in the AB-system
unless otherwise stated. 

\section{Photometric Redshift Technique}

The photometric redshift code developed for COSMOS is based on template
fitting technique (Gwyn 1995; Mobasher \etal\   1996; Chen \etal\   1999; 
Arnouts \etal\  ., 1999; Benitez 2001; Bolzonella \etal\   2000; ). The
templates, representing the rest-frame Spectral Energy Distribution (SED) 
for galaxies of different types, are convolved with the
response functions of filters used in the COSMOS photometric 
observations. 
These were then shifted in redshift space and fitted to the observed SEDs 
of individual galaxies by minimizing the $\chi^2$ function, 

$$\chi^2 = \Sigma_{i=1}^n ((F^i_{obs} - \alpha F^i_{template})/\sigma^i)^2$$

\noindent The summation is over the passbands (i.e. number of 
photometric points) and $n$ is the total number of passbands. $F^i_{obs}$ and
$F^i_{template}$ are, respectively, the observed and template 
fluxes for each passband; $\sigma^i$ is the uncertainty in the observed
flux and $\alpha$ is the overall flux normalisation. The redshift 
corresponding to the centroid of redshift probability distribution and 
SED (i.e. spectral type) yielding the minimum $\chi^2$ value are  
then assigned to each galaxy. The redshift 
probability function for each galaxy is defined  
as $p(z,T)=e^{-\chi(z,T)^2/2}$, where $z$ and $T$ are respectively, 
the redshift and spectral type of galaxies. The estimated redshift 
corresponds to the centroid of this probability distribution, defined as

$$z= \frac{\int^{T_{max}}_{T_{min}}\int^{z_{max}}_{z_{min}} p(z,T) z dz dT
}{\int^{T_{max}}_{T_{min}}\int^{z_{max}}_{z_{min}} p(z,T)dz dT}$$

\noindent This is used as the best estimate for the photometric redshifts 
in this study. 
The code gives the option of using Bayesian priors based on luminosity functions (LFs). 
The main effect of a LF prior is to discriminate between cases in which 
the redshift probability distribution shows multiple peaks 
due to ambiguity 
between the Lyman and 4000 \AA\  features. 
The inferred absolute magnitudes of the galaxy if it is at either of the 
redshift peaks  
can then be used to discriminate between these possibilities (i.e. 
an implied absolute magnitude
significantly brighter than $M^\ast$ is increasingly unlikely). 
Thus, for each 
redshift, we calculate the rest-frame absolute V-band magnitude  
and compare it to the LF. For this study we use a Schecter LF
with $M^\ast =-22 $ mag and faint-end slope $\alpha=-1.26$.  
This corresponds to the mean of the characteristic magnitudes
and faint-end slopes of the B-band luminosity functions for all spectral
types of galaxies and over the redshift range $0 < z < 1$ 
(Dahlen \etal\ 2005), converted to V-band absolute magnitude using
rest-frame $B-V$ colors. Compared to B-band, the V-band luminosity function
(LF) is less sensitive to details of the spectral types of galaxies, 
allowing us
to use a single LF for all types. In any case, the final
photometric redshifts are not dependent on the LF used.  
Evolution with redshift of both $M_V^\ast$ and faint-end slope of the LF 
(Dahlen et al 2005) are incorporated into the LF prior.  
Nevertheless, we explored sensitivity of our results on different
choices of $M^\ast$ and $\alpha$ and found them to be relatively insensitive
to the choice of these parameters. Finally, using the spectroscopic 
sample (section 4.3), we optimised the prior LF parameters to minimise
the scatter between the estimated
photometric and spectroscopic redshifts.

We also include internal extinction ($E_{B-V}$) as a free parameter
in the $\chi^2$ minimisation process (alongside redshift and spectral types)
and estimate $E_{B-V}$ for individual galaxies using 
Galactic extinction law for early-type galaxies and 
Calzetti law (Calzetti \etal\   2000) for late-type and starbursts. 
Absorption due to 
intergalactic HI is included using the parametrization in Madau (1995). 

Basic template spectral energy distributions (SEDs) for normal
galaxies (E, Sbc, Scd and Im) from Colman \etal\   (1980) and
two starburst templates are from Kinney \etal\   (1996)-(SB2 and SB3- Figure 1).
  The templates are
corrected for systematic calibration errors and extended to the ultraviolet
and infrared wavelengths using the method of Budavari et. al. (2000).
The template corrections were derived from over 3000 galaxies with 
spectroscopic
redshifts in the Hawaii Hubble Deep Field North (H-HDFN) (Capak et. al.
2004; Cowie  \etal\  2004; Wirth \etal\  2004; Treu \etal\  2005; Steidel \etal\  2004; Erb \etal\  2004). These galaxies had deep optical and infrared 
photometry
(U,B$_J$,V$_J$,R$_c$,I$_c$,z$^+$,J,H,K$_s$,HK$^\prime$)- (Capak \etal\
2004; Bundy \etal\ 2005; Wang \etal\ 2005). Our corrections
in the optical and UV are consistent with the calibration errors estimated by
Coleman \etal\  (1980) and Kinney \etal\  (1996).  The largest correction is
in the UV where Coleman \etal\ (1980) forced agreement between their 
ground-based and IUE data.  The infrared properties of our templates do 
differ significantly
from those extended using Bruzual \& Charlot (2000) models (Bolzonella \etal\  2000).
 This is not surprising since stellar population models have
large uncertainties in the infrared (Maraston 2005).  The details of our
template optimisation method will be discussed in Capak \etal\ (in prep 2007).
The final, modified template SEDs are shown in Figure 1. we constructed 
intermediate-type templates from 
the weighted mean of the adjacent templates,    
defining five intermediate-type templates between the main
spectral types. Our fitting therefore included a total of 31 SED templates, 
each redshifted between $z=0$~and $z=6$~in $\Delta z=0.01$ steps.

\section{Photometric Data }
 
The photometric observations for COSMOS were carried out at
optical ($u^\ast$: CFHT; B{\it g}V{\it riz:} SuprimeCam/Subaru; $i-$: CFHT; 
$i_{814}$: ACS/HST) 
and near-Infrared ($K_s$: Flamingos/CTIO and Kitt Peak) wavelengths. 
We also obtained
narrow-band survey of the COSMOS field at 815 nm (NB815 filter) with
SupremeCam/Subaru.  
The response functions for these filters are shown in Figure 2.  
Table 1 lists the filters, including
the effective wavelength and band-width for each filter and the 
corresponding depth and image seeing. Details of the ground-based observations 
and data reduction are presented in Capak \etal\   (2006) and Taniguchi \etal\   
(2006). 

The reduced images in all bands were PSF matched by Gaussian convolution  
with FWHM corresponding to the worst 
seeing (1.5'' in $K_s$ band), allowing for non-Gaussian wings of the PSFs.   
The multi-waveband photometry catalog was then generated using 
SExtractor (Bertin \& Arnout 1995). This is first done by measuring the total
($mag_{auto}$) and aperture (3'' diameter) magnitudes on the detection
image ($i-$band) and, for each galaxy, estimate the correction from
aperture to total magnitudes. This correction is subsequently applied
to the respective galaxies, detected in other bands. Details of 
the photometry, star/galaxy separation and catalog generation are given 
in Capak \etal\   (2006). 

In the next section we simulate the COSMOS catalog by constructing a similar
mock galaxy catalog with the same filters, depths and SED shapes and assign a
random redshift to each simulated galaxy. The simulated catalog will then 
be used to test the accuracy of our estimated photometric redshifts and the
consistency of our technique. This
will be further examined by comparing the photometric and spectroscopic 
redshifts to a sample of COSMOS galaxies with available such data.

%We match the photometric catalog with the
%A total of ** galaxies in the COSMOS field have spectroscopic redshifts 
%from 2-degree Field Galaxy Redshift Survey (2dFGRS; ** galaxies) and 
%VIRMOS (VVDS; ** galaxies). We matched the multi-waveband photometric catalog
%with the spectroscopic data and constructed a spectroscopic catalog containing
%a total of ** sources, covering the redshift range $0 < z < 2$ with 
%$i_{AB} < 24$. The spectroscopic catalog will be used to examine the 
%accuracy of photometric redshifts and to calibrate the photometric redshift 
%technique. 

\section{Simulations}

\subsection{Mock Catalog}

To explore the accuracy of photometric redshifts, 
we generated mock catalogs consisting of galaxies with the SEDs shown in 
Figure 1
and photometry measured in the same filters used for COSMOS (Figure 2). 
The aims of the simulation is 
to explore dependence of the photometric redshifts on 
the $S/N$ ratio, magnitude limit, redshift and galaxy type and how we could
minimise the number of outliers (objects with very different output and input 
photometric redshifts). 

We use the rest-frame B-band LFs derived for different spectral types of
galaxies, using the GOODS data (Dahlen \etal\   2005);  therefore, both the
type-dependence and evolution of the LFs are incorporated
into the simulations. 
Each galaxy is assigned a random absolute magnitude in the range
$-24 < M_B < -16$ mag and spectral type, drawn from the type-dependent LFs. 
The six galaxy templates are the same as used
in the photometric redshift calculation (\S2). To each simulated galaxy, 
we also specified a redshift in the range $0 < z < 6$. 

For any given galaxy, the K-correction in each band was estimated by 
convolution of the
filter responses with the SED associated with that galaxy, 
shifted to its assigned redshift. We then estimate the  
apparent magnitudes  using the rest-frame absolute magnitudes and 
the distance moduli.   
We restrict the mock catalog to galaxies with apparent magnitudes (in any
given band) brighter than the observed magnitude limits in COSMOS (Table 1).  
We also assign photometric errors to each magnitude,  
depending on the S/N ratios with the same depth as real COSMOS data. 
Ideally, the photometric errors need to be
estimated, also taking into account blending or defects on the images
(sources near bright objects and internal camera reflections). This can be
performed by distributing images of galaxies with known  
mag/type/redshift into the existing multi-band images, apply the same 
photometric measurement scheme and then running the photometric redshift 
code. Therefore, simulations here only provide an internal consistency check 
in reproducing the input parameters. 

The resulting catalog consists of simulated data, including: magnitudes and 
their associated errors in the same filters as in the initial COSMOS catalog, 
redshifts and spectral types for each galaxy, all consistently derived. The
mock catalog contains a total of $\sim$96,000 galaxies 
to a magnitude limit of $i=26.2$ mag (S/N=5 - similar to the observed 
COSMOS catalog). 

%In Table 2, we list
%changes in the surface density of galaxies as a function of $i-$band 
%magnitude limit, only including objects with S/N ratios higher than 5, 10, 
%20 and 40. 

The photometric redshift code (\S2) was used to estimate redshifts and
spectral types for mock galaxies, using prior and considering extinction 
as a free parameter. Results from the simulated
catalog, showing the performance of the code, are presented in Table 2 where, 
for each value of the magnitude limit (i.e. $S/N$ ratio) 
and spectral type, we estimate the {\it rms} scatter in photometric 
redshift error, defined as; 
$\Delta(z) = (|z_{output} - z_{input}|)/(1 + z_{input})$, the fraction and 
total
number of outliers, defined as galaxies with 
$\Delta(z)> 3\sigma (\Delta(z))$, ~and 
changes in the median redshift as a function of the S/N ratios. 
The simulation results in Table 2 clearly illustrates that the accuracy 
of photometric 
redshifts decreases as the limiting magnitude becomes fainter and the 
S/N ratio is reduced. Moreover, we find that for early-type 
galaxies there is better
agreement between the input and output redshifts, with a smaller
fraction of outliers, compared to late-type and starbursts. This is likely
due to a stronger 4000\ \AA break in ellipticals compared to later type
galaxies. 

Figure 3 shows comparison between the input and output redshifts as a 
function of i-band magnitude and $S/N$. At $ i > 25$ mag, the photometric 
redshift accuracy
starts to significantly degrade. It is clear that at higher $S/N$
values (i.e. brighter $m_i$), photometric redshift code recovers the input 
redshifts. Also, most of the scatter at faint magnitudes (low S/N)
is due to late-type galaxies and starbursts. 
This will be used as a guide to adopt the photometric or magnitude
limit of the sample in order to optimise photometric redshift measurement. 

\subsection{Accuracy of photometric redshifts}

The simulation results can be used to define a useful measure of the 
photometric redshift accuracy for each galaxy. This parameter
is defined as 

$$D_{95} = {\Delta_{95}\over (1+z_{output})}$$

\noindent where $\Delta_{95}$ is the 95\% confidence interval (i.e. the width of
the redshift probability distribution corresponding to 95\% confidence
interval) and $z_{output}$ is the estimated photometric redshift. Therefore, 
 $D_{95}$ can be calculated independent from any knowledge about spectroscopic 
redshift. If the error distribution
is Gaussian, then, by definition, $\Delta_{95} = 2\sigma_z$. 

To explore how $D_{95}$ is related to the accuracy of photometric 
redshifts, we study the correlation between $D_{95}$ with 
$\sigma (\Delta(z))$ and $\Delta(z)$, using the
mock cataloge, as shown in Table 3 and Figure 4 respectively.  
The sample used in Table 3 is limited to galaxies with  $S/N>10$.  
This is to minimise photometric uncertainties and to uncouple performance of
different photometric redshift error estimators independent from 
photometric problems at faint flux levels. 
For simulated sub-samples, selected based on $D_{95}$ limits 
(Table 3; column 1), we estimate $\sigma (\Delta(z))$ values 
for the full sample and when excluding the outliers, defined as
galaxies with $\Delta(z) > 3 \sigma (\Delta(z))$.  Results are listed
in Table 3, where it  
shows a clear decrease in $\sigma (\Delta(z))$ values and in fraction of the 
outliers towards smaller $D_{95}$. This demonstrates that
$D_{95}$ provides a useful and practical measure to identify the fraction of
outliers. Moreover, 
the median redshift of the survey is found to be independent of $D_{95}$, 
due to our $S/N$ cut. 

The scatter in $\Delta(z)$ increases with increasing $D_{95}$  
and for fainter magnitude limits. For galaxies with  $D_{95} > 0.2$, the
scatter in $\Delta(z)$ significantly increases, indicating an increase
in photometric redshift errors. For fainter galaxies 
($i > 24.2$ mag), where the accuracy of photometric redshifts decreases, 
we find an increase in $D_{95}$ parameter and larger scatter in $\Delta(z)$. 

In summary, $D_{95}$ enables the identification of outliers in derived 
photometric redshifts, independent at all redshifts in the sample. 

\subsection {Comparison with Spectroscopic Redshifts}

The ultimate test of the accuracy of photometric redshifts is the
comparison with the spectroscopic redshifts. The spectroscopic sample
here consists of galaxies observed to $i_{AB}\sim 24$ mag. in 
the $z$COSMOS program, using 
VIMOS on VLT (Lilly \etal\   2006). We select 958 galaxies 
with the most reliable spectroscopic redshifts (based on two or three lines). 
We restrict the sample to redshift range $z < 1.2$, as beyond this,  
the 4000 \AA\ break lies at the edge of the optical bands. Also, due to the 
relatively shallow depth of our $K_s$-band data, these are not available
for fainter galaxies. This reduces total number of galaxies with spectroscopic
redshifts to 879. 

Photometric redshifts were derived using the techniques described in 
\S3 and compared with the spectroscopic redshifts in Figure 5.  
The effects of the luminosity function prior and extinction corrections 
are also explored. A total of 12
galaxies in the spectroscopic sample ($z < 1.2$) were identified as AGNs 
from their X-ray emission (Brusa \etal\   2006). 
The AGNs were removed from 
the spectroscopic sample and only
the ``normal'' galaxies were used in the comparison.

We measure the $D_{95}$ parameter for the 868 galaxies with $z < 1.2$ 
in our spectroscopic sample. The relation between $D_{95}$ and 
$\Delta(z)=(z_{phot}-z_{spec})/(1+z_{spec})$ is shown in Figure 6a. 
Galaxies with $D_{95} < 0.2$ are seen to have, on average, 
$\Delta(z)\sim 0$ although with some scatter. This confirms that, on average, 
$D_{95}$ parameter provides a good measure of the reliability of photometric 
redshifts. Distribution of $D_{95}$ 
values for three spectral types of galaxies (elliptical, spiral and 
starbursts) in the spectroscopic sample are presented in Figure 6b. The
width of the distributions for
different types are consistent with the observed scatter in Figure 6a. 
The peak of the $D_{95}$ distributions are at $D_{95} \sim 0.08$ (for
ellipticals) and 0.12 (for spirals and starbursts), indicating the
reliability with which one could measure photometric redshifts for
different spectral types of galaxies. 

Table 4 compares the $\sigma (\Delta (z)) $ values and the fraction of 
outliers, defined as objects with $ \Delta (z) > 3 \sigma (\Delta (z))$, 
for different cases
(with and without prior and extinction). It is clear from Table 4 and
Figure 5 that the best agreement between the photometric and spectroscopic
redshifts are obtained when both prior and extinction corrections 
are enabled. In its best case, this corresponds to an {\it rms}
of $\sigma (\Delta(z)) =0.031$. This is consistent with the {\it rms} 
estimated from the simulations in \S3.  
It is also clear from Table 4 that $D_{95}$ parameter is directly
correlated with the fraction of outliers, as defined by 
$\sigma (\Delta (z)) $- (i.e. deviation of photometric
redshift from its spectroscopic counterpart). 
No trend is found between redshift and spectral 
types in Figure 5, indicating there is no significant bias in redshift 
estimates as a function of spectral type. 

Finally, the relation between $\Delta(z)$ and i-band magnitudes for 
``normal'' galaxies in the spectroscopic sample is shown in Figure 7.  
The errors in the photometric redshift shows no
dependance on the magnitude of the galaxies or their spectral type.  

We divide galaxies into spectral type bins (ellipticals, spirals and
starbursts) and compare their estimated photometric and spectroscopic
redshifts, as listed in Table 4. The photometric redshifts are
estimated for the case assuming prior and extinction (the optimum case),  
considering all galaxies regardless of their 
$D_{95}$. We find comparable $\sigma (\Delta(z))$ values for elliptical 
(0.034), spiral (0.030) and starbursts (0.042). For each of the scenarios
in Table 4, we also estimate the fraction of galaxies (with respect to total) 
of different spectral types. The result, listed in Table 4, shows a 
simultaneous decrease in the fraction of ellipticals and increase in
the fraction of starbursts when extinction correction is enabled. 
No significant change in the fraction of spirals is observed. 

Figure 5 shows a reduction in $\sigma (\Delta (z))$ for ellipticals
when extinction correction is applied, with this having a less significant 
effect for the starbursts, contrary to expectations. 
However, as shown 
in Table 4, we find a change in the fraction of both ellipticals and
starbursts when dust extinction is included as a free parameter in the
photometric redshift fits. This indicates a change in the 
best-fit spectral types of
galaxies (Figure 5), depending wether or not we apply the dust extinction 
correction. The derived spectral types of elliptical and later type 
(spirals, irregulars and starbursts) galaxies here are examined by
comparing them with independently estimated quantitative morphologies 
(compactness, asymmetry and Gini coefficients). These morphological
parameters are consistent with the derived spectral types 
(Capak \etal\   2006). 

We now estimate photometric redshifts for 12 AGNs with $z < 1.2$, 
including the prior and extinction. These are compared with
their spectroscopic redshifts in Figure 8 and show that the {\it rms} scatter 
is again lowest when including the prior and correcting for
local extinction.  This corresponds to 
$\sigma (\Delta(z)) = 0.10$ (Table 4). 
The small {\it rms} measured for AGNs (type II) indicates that once 
extinction fitting is enabled, one can derive their photometric
redshifts using templates based on normal galaxies. 

\section {Other Photometric Redshift Codes}

In this section we explore  how  photometric redshifts depend  on different
techniques, codes and choice of priors, using a variety of photometric
redshift codes. We compare results from the code presented in the previous 
section (refered to as "COSMOS") with three other codes: 
Zurich  Extragalacitc Bayesian Redshift Analyzer 
(ZEBRA; Feldmann \etal\ 2006), Le Phare (Arnout  1999) and Baysian Photometric 
Redshift code (BPZ; Benitez 2000). Here we give a summary of basic 
characteristics of these codes.

\noindent {\bf The Zurich Extragalactic Baysian Redshift Analyzer (ZEBRA):} ZEBRA 
estimates redshifts and template types of galaxies using medium- and 
broad-band photometric data (Feldmann \etal\   2006). In the photometry check 
mode, for each galaxy
and in any given filter, ZEBRA computes the difference between the observed 
magnitudes and those predicted by templates, using a training set
with available spectroscopic redshifts. A linear (or higher order) regression
is then applied to the relation between the residual and observed galaxy
magnitude, with a constant offset estimated and subsequently applied to 
magnitudes in any given filter. In the template check mode, ZEBRA uses the
$\chi^2$ minimization technique to optimize the difference between the 
observed and template-based fluxes for all passbands, averaged over
all galaxies in the photometric catalog. By introducing additional terms
to the $\chi^2$ equation, ZEBRA prevents too large deviations between the
observed and model templates and regularizes the template shapes. It is run
in both Maximum Likelihood and Baysian modes. In the later case, a prior is 
calculated in redshift and template space, using an iterative procedure.
%ZEBRA estimates photometric redshifts to an accuracy of   ($\Delta z/(1+z_s)=0.054$
In the current release of this code, reddening due to dust extinction 
is not included.
 
\noindent {\bf Le Phare photometric redshift code:} 
The Le Phare code (Arnouts \etal\ 1999) is based on $\chi^2$ fitting 
method, comparing the observed magnitudes with those predicted from an 
SED library. This simultaneously runs libraries for stars, galaxies 
and quasars, which are then used to separate different classes of objects. 
An automatic calibration method is applied by using the spectroscopic 
redshift sample as training set (Ilbert \etal\  ., 2006). This adaptive method 
combines an iterative correction of the photometric zero-points and an 
optimisation of the SED templates. It allows to remove systematic differences 
between the spectroscopic and photometric redshifts and reduce the fraction of
catastrophic failures. Reddening correction is applied to templates later 
than Sbc types,  using the small Magellanic cloud extinction law. 
In this work, we adopt the same empirical templates as Ilbert et
al. (2006).  An additional Bayesian approach has been used, involving 
 priors based on redshift distributions, following the formalism of
 Benitez (2000).  
 %and an accuracy of $\sigma(\Delta z/(1+z_s)=0.031$ with $\eta=1.0\%$ of catastrophic failures (defined as $\Delta z/(1+z_s)>0.15$.

\noindent {\bf Baysian Photometric Redshift Code (BPZ):} The Bayesian approach considers
the redshift distribution, $p(z|C,m)$, as a function of the observed color 
($C$) and magnitude ($m$)- (Benitez 2000). The prior used here is therefore
based on the probability of a galaxy having redshift, $z$, and spectral type,
$T$, given its magnitude. This is different from a luminosity function based
prior used in the previous sections (the COSMOS code). Therefore the BPZ code 
provides redshifts based on both maximum liklihood and prior based 
techniques. 
The prior-based photometric redshifts from the BPZ are generally found to be 
more accurate than the results obtained  when no priors are used. 
%In its application to this work, BPZ provides photometric redshifts with an accuracy of $\sigma(\Delta
 %z/(1+z_s)=0.028$ with $\eta=1.0\%$ of catastrophic failures (defined
 %as $\Delta z/(1+z_s)>0.15$. .

The four codes are not completely identical and hence, we need to
specify any intrinsic differences between them when comparing results from 
the codes. We present a list of the set up parameters used in each of the 
above codes in  Table 5.

\subsection {Comparison between different photometric redshift codes}

The four photometric redshift codes have been applied on the 
{\it same} spectroscopic sample, with the
$\Delta z = (z_{phot} -z_{spec})/(1+z_{spec}$) distributions
compared in Figures 9 (without prior) and 10 (with prior). The
$\Delta z$ distributions from the codes used here are approximately fitted
by a  Gaussian with $\sigma = 0.026$ (Figures 9 and 10). However, 
the distributions for some codes are slightly offset from $\Delta z =0$, with
extended wings. 

 The absolute accuracy in each code depends on the way the outliers are
defined. To directly compare the photometric redshift accuracy from 
various codes, we follow the same procedure for all the four photometric 
redshift codes and present the results in Table 6 (with no priors) and 
Table 7 (with priors). 
For each code, we calculate the upper and lower 68\% intervals (top-left and top-right number in each grid) from the distribution of $\Delta(z)$ between the photometric and
spectroscopic redshifts and between the photometric redshifts from
different codes.  This is a different definition than the ``average'' {\it rms}
values presented for COSMOS photometric redshifts in Table 4 and is 
defined to more clearly show the asymmetry in $\Delta (z)$ distributions
between different codes. This also explains the difference in the scatter
between photometric and spectroscopic redshifts found here (Tables 6 and 7)
compared to that listed in Table 4. 

Assuming a Gaussian distribution
for $\Delta(z)$ values, that would correspond to 1$\sigma$ standard deviation and for symmetric distributions the top-left and the top-right number should be the same. 
 Objects with $\Delta(z)$ values outside the 1$\sigma$ limit (bold number in each grid in Tables 6 and 7) are considered as outliers. This prescription defines the accuracy independent of the definition of 
the outliers. 

The comparison between the estimated redshifts from various photometric 
redshift codes  with their spectroscopic 
counterparts are also shown  on the first row of Tables 6 and 7.
The rest of the entries present comparison between the different codes.  
Results listed in the tables show excellent
agreement between different photometric redshift codes, with all agreeing 
well with the spectroscopic redshifts. However, there is a slight
improvement in the {\it rms} scatter for COSMOS code when using the
prior while, prior has no such effect on other codes. This is likely
due to the fact that the prior here was partly optimised on the spectroscopic
data, using the photometric data set. 
 
%To explore if the deviant objects are
%the same sources with failed photometric redshifts from different codes, 
%we also compute the {\it rms} values between the photometric redshifts 
%from these codes

\section{Analysis of Photometric data}

In the previous sections we demonstrated that one could derive reliable
photometric redshifts, using the available multi-waveband data for 
galaxies in COSMOS. These are extensively used in the analysis of 
COSMOS dataset. In this section we present preliminary results, using
the photometric redshifts for the entire COSMOS galaxies with $i< 25$ mag. 
Given the results in Table 4, we use prior and consider extinction as
an independent parameter in the fit.

The photometric redshift distributions for different spectral types of 
galaxies
in COSMOS are presented in Figure 11. Only galaxies with $i < 25$ 
mag are used here, as they have the most reliable photometric redshifts. 
Moreover, as discussed in \S4.3, we restrict the sample to galaxies 
with $z < 1.2$. 
There is similar distribution for all the spectral types with 
redshift. The photometric redshift distribution for COSMOS
(to $i_{AB} < 24$) is compared in Figure 12 with the spectroscopic redshift
distribution for the VVDS to the same depth (Le Fevre et al. 2005), after
normalising the number of sources to the areas of their respective surveys. 
The overall agreement is good, with similar median redshifts. The VVDS
only targets 25\% of the galaxies to its spectroscopic magnitude limit. 
This, combined with the difficulty in measuring spectroscopic redshifts 
for fainter galaxies in VVDS and cosmic variance are responsible for the
observed difference between the two distributions in Figure 12.  

In Figure 13 we present rest-frame absolute magnitudes ($M_V$) for COSMOS 
galaxies. These are estimated using 
its best-fit photometric redshift and spectral type, following the prescription
described in Dahlen \etal\   (2005). As expected, there is a trend in
$M_V$ absolute magnitudes with spectral types, with objects with earlier 
types being brighter. The median absolute magnitudes correspond to
$M_V = -21.3 $ (E/SO), $-20.5$ (Sa/Sb), $-19.7$ (Sc), $-18.7$ (starbursts). 
 
\subsection{Stellar Mass Estimates}

The stellar mass for COSMOS galaxies is measured using the relation between
$M/L_V$ and rest-frame $(B-V)_0$ colors
$$M/L_V = -0.628 + 1.305\ (B-V)_0$$
Bell \etal\   (2005). We assume Salpeter IMF with 
$0.1 M_\odot < M < 100 M_\odot$. Average rest-frame $<B-V>_0$ colors, 
corrected for extinction, 
are estimated for each spectral type (E, Sa, Sb, Sc, Im and starburst), using
the appropriate templates. Then, to each galaxy, using its
best-fit spectral type (which is derived consistently with its estimated
extinction and photometric redshift), we assign the $<B-V>_0$ color and 
hence, the $M/L_V$ ratio from the above equation. 
Combined with rest-frame absolute V-band magnitudes ($M_V$), the stellar 
mass is then estimated as
$$log (M_{stellar}/M_\odot) = M/L_V - 0.4\ (M_V - 4.82)$$

\noindent K-band luminosities, being produced by evolved stellar population in
galaxies, are more directly correlated with the stellar mass in galaxies. 
However, due to shallowness of our K-band data over the COSMOS area, many
galaxies are not detected in this band. Therefore, we use the V-band
luminosity as a proxy for the K-band to measure the stellar mass. For 
a sub-set of our galaxies, the stellar masses measured using the K- and 
V- band luminosities were compared and agree better than 5\%. However, 
by definition, this is a sample dominated by the most massive and reddest
galaxies and therefore, this cannot be used as a measure of the accuracy
for stellar masses for the rest of the galaxies in this sample. The
main source of uncertainty in our stellar mass estimates here is 
the scatter in the mean $<B-V>_0$ colors for each spectral type and
the accuracy with which the spectral types are measured for individual
galaxies. 

In Figure 14 we present the distribution of $M_{stellar}/M_\odot$ values 
as a function of spectral type and redshift. In a given redshift range, 
elliptical and early-type spiral galaxies are more massive than later type
galaxies. However, for a given spectral type of galaxies, we find an
increase in galaxy mass with redshift. This is likely caused by a bias
in our magnitude limited sample, due to selecting brighter galaxies at
higher redshifts. 

\section{Summary}

We develop a photometric redshift code and use that to measure redshifts
and spectral types for galaxies in the COSMOS survey. The technique uses
$\chi^2$ template fitting, combined with luminosity function priors and
with the option to estimate internal extinction ($E(B-V)$). We use 
extensive simulations to examine reliability of the code and study 
its accuracy as a function of photometric magnitude limits and $S/N$ ratios. 
We define a new parameter, $D_{95}$, to identify the objects with 
catastrophic failure in photometric redshift estimate. 

We estimate photometric redshifts for a sample of 868 galaxies with 
available spectroscopic redshifts (to $z < 1.2$) from $z$COSMOS. 
Considering different scenarios, dependeing on using prior and/or
extinction, we compare the photometric and spectroscopic redshifts 
for this sample. The best agreement is found when invoking both
prior and dust extinction correction, giving $\sigma (\Delta(z)) = 0.031$, 
where $\Delta(z) = (z_{phot} - z_{spec})/(1 + z_{spec})$. This gives
a small fraction of outliers (2.5\%). For a sample of 12 type II AGNs
with available spectroscopic redshifts, we
find $\sigma (\Delta(z)) = 0.10$. 

Our photometric redshift code here is compared with three independent codes. 
The estimated redshifts are in excellent agreement. We measure 
photometric redshifts and spectral types for the entire COSMOS galaxies
and present preliminary results concerning redshift and absolute magnitude
distributions. We use 
the estimated photometric redshifts and spectral types to measure
stellar masses of galaxies and study changes in stellar mass
among galaxies with different spectral types and with redshift.

\clearpage

\begin{deluxetable}{ccccccc}
\tablecaption{Data Quality and Depth\label{t:data-depth}}
\tablehead{
\colhead{Filter}& \colhead{Central}         & \colhead{Filter}   &
\colhead{Seeing} & \colhead{Depth\tablenotemark{1,2}}  &
\colhead{Saturation\tablenotemark{2}} & \colhead{Offset from
\tablenotemark{3}}\\
\colhead{Name}         & \colhead{Wavelength (\AA)} & \colhead{Width (\AA)}
& \colhead{Range (\asec)}  & \colhead{}             & \colhead{Magnitude} &
\colhead{Vega System}}
\startdata
$u^\prime$    &    3591.3    &    550        &    1.2-2.0    & 22.0 & 12.0 &
0.921\\
$u^*$        &    3797.9    &    720        &    0.9        & 26.4 & 15.8 &
0.380\\
$B_J$        &    4459.7    &    897        &    0.4-0.9    & 27.3 & 18.7 &
-0.131\\
$g^\prime$    &    4723.1    &    1300    &    1.2-1.7    & 22.2 & 12.0 &
-0.117\\
$g^+$        &    4779.6    &    1265    &    0.7-2.1    & 27.0 & 18.2 &
-0.117\\
$V_J$        &    5483.8    &    946        &    0.5-1.6    & 26.6 & 18.7 &
-0.004\\
$r^\prime$    &    6213.0    &    1200    &    1.0-1.7    & 22.2 & 12.0 &
0.142\\
$r^+$        &    6295.1    &    1382    &    0.4-1.0    & 26.8 & 18.7 &
0.125\\
$i^\prime$    &    7522.5    &    1300    &    0.9-1.7    & 21.3 & 12.0 &
0.355\\
$i^+$        &    7640.8    &    1497    &    0.4-0.9    & 26.2 &
20.0\tablenotemark{*} & 0.379\\
$i^*$        &    7683.6    &    1380    &    0.94        & 24.0 & 16.0 &
0.380\\
$F814W$        &    8037.2    &    1862    &    0.12        &
24.9\tablenotemark{+} & 18.7 & 0.414\\
$NB816$        &    8151.0    &    117        &    0.4-1.7    & 25.7 & 16.9
& 0.458\\
$z^\prime$    &    8855.0    &    1000    &    1-1.7    & 20.5 & 12.0 &
0.538\\
$z^+$        &    9036.9    &    856        &    0.5-1.1    & 25.2 & 18.7 &
0.547\\
$K_s$        &    21537.2    &    3120    &    1.3        & 21.6 & 10.0 &
1.852\\
\enddata
\tablenotetext{1}{$5\sigma$ in a 3\asec aperture.}
\tablenotetext{2}{In AB magnitudes.}
\tablenotetext{3}{AB magnitude = Vega Magnitude + Offset.  This offset does
not include the color conversions to the Johnson-Cousins system used by
Landolt (1992).}
\tablenotetext{*}{Compact objects saturate at $i^+<21.8$ due to the
exceptional seeing.}
\tablenotetext{+}{The sensitivity for photometry of a point source in a
0.15\asec aperture is 26.6,  for optimal photometry of a 1\asec galaxy it is
26.1}
\end{deluxetable}

\begin{table*}

\caption[]{Photometric redshift accuracy from the COSMOS simulations for different limiting magnitudes. Outliers here are defined as objects with $\Delta(z) > 3 \sigma (\Delta (z)) $, where $\Delta(z) = (|z_{output} - z_{input}|)/(1 + z_{input})$.}
\begin{tabular}{llcccc}
& & & & & \\
$m_{lim}$ & $\sigma (\Delta(z))$ & $\sigma (\Delta(z))$ & Fraction of & Median z & Fraction types\\
 &   full sample & w/o outliers & outliers (\%) & & (\%)\\
& & & &  & \\
$ < 26.2$ & 0.183 & 0.140 & 2.1 & 1.13 & \\
$ < 25.7$ & 0.142 & 0.088 & 2.2 & 0.96 & \\
$ < 25.2$ & 0.089 & 0.048 & 1.1 & 0.81 &\\ 
$ < 24.7$ & 0.054 & 0.031 & 0.62 & 0.73 &\\
$ < 24.2$ & 0.042 & 0.025 & 0.37 & 0.67 &\\
& & & & & \\
$i < 25.2$: & & & & &  \\
early-type & 0.058 & 0.031 & 0.40 & 0.90 & 14\\
late-type  & 0.085 & 0.045 & 0.83 & 0.78 & 60\\
starburst & 0.11 & 0.065 & 1.1 & 0.84 & 26\\
\end{tabular} 
\end{table*}

\begin{table*}

\caption[]{Relation between $D_{95}$ and the photometric redshift accuracy
($\sigma (\Delta(z))$) from the COSMOS simulations only using objects
with S/N $>$~10.
The outliers here are defined as objects with
$ \Delta(z) > 3 \sigma (\Delta(z))$ and are measured for the samples
selected based on $D_{95} > D_{95}^0$, where
$D_{95}^0$ values are listed in column 1.}
\begin{tabular}{lcccccc}
& & & &   \\
$D_{95}$ & Spectral& $\sigma (\Delta(z))$ & $\sigma (\Delta(z))$ & 
Fraction of & Median z & Fraction of \\
  &types  & full sample & w/o outliers & outliers (\%) & & objects 
(\%)  \\
& & & & & & \\
all objects & all & 0.114 & 0.066 & 1.5  & 0.91 & 100  \\
 &early & 0.061 & 0.034 & 0.57 & 0.93 &  12 \\
 &late & 0.11 & 0.062 & 1.6 & 0.84 &  60 \\
 &starburst & 0.14 & 0.084 & 1.5 & 0.92 &  28 \\
$< 0.7$ &all& 0.056 & 0.042 & 2.1  & 0.96 & 83 \\
 &early & 0.034 & 0.028 & 1.8  & 0.95 &  14 \\
 &late & 0.053 & 0.040 & 1.9 & 0.86 &  59 \\
 &starburst & 0.072 & 0.055 & 2.5 & 0.92 &  27 \\
$< 0.5$ &all& 0.041 & 0.033 & 1.7  & 0.93 & 72 \\
 &early & 0.034 & 0.028 & 1.8  & 0.94 &  16 \\
 &late & 0.038 & 0.032 & 1.8 & 0.81 &  60 \\
 &starburst & 0.049 & 0.039 & 1.5 & 0.84 &  24 \\
$< 0.3$ &all& 0.030 & 0.026 & 0.80  & 0.82 & 58\\
 &early & 0.027 & 0.024 & 1.3  & 0.89 &  18 \\
 &late & 0.027 & 0.025 & 1.1 & 0.72 &  60 \\
 &starburst & 0.038 & 0.028 & 2.5 & 0.65 &  22 \\
\end{tabular} \label{table1}
\end{table*}

\begin{table*}

\caption[]{\bf Comparison with spectroscopic redshifts. Outliers here are defined the same as in Table 3}
\begin{tabular}{lccccccc}
& & & & & & &   \\
  $D_{95}$ & $\sigma (\Delta(z))$\tablenotemark{1} & $N_{tot}$\tablenotemark{2} & $N_{outlier}$\tablenotemark{3} &Outlier fraction & $n_E$ & $n_{sp}$ & $n_{Starburst}$\\
%& & & &    \\

\multicolumn{8}{l}{\bf no prior + no extinction corr.}\\
%& & & &    \\
all     & 0.091 (0.042) & 868 & 5  &  0.006  &  0.25 & 0.63 & 0.12 \\
$< 0.2$ & 0.047 (0.035) & 828 & 18 & 0.022 & & & \\
$< 0.3$ & 0.047 (0.035) & 841 & 19 & 0.023 & & & \\
%& & & &    \\
\multicolumn{8}{l}{\bf no prior + extinction corr.}\\
%& & & &    \\
all     & 0.086 (0.034) & 868 & 4  & 0.005 & 0.20 & 0.52 &  0.28 \\ 
$< 0.2$ & 0.036 (0.029) & 779 & 15 & 0.019 & & & \\
$< 0.3$ & 0.036 (0.029) & 830 & 15 & 0.018 & & & \\
%& & & &    \\
\multicolumn{8}{l}{\bf with prior + no extinction corr.}\\
%& & & &    \\
all      & 0.17 (0.047) & 868  &  5  & 0.006  & 0.24 & 0.65 & 0.11 \\
$ < 0.2$ & 0.044 (0.033) & 841 & 18 & 0.021  & & &\\
$ < 0.3$ & 0.045 (0.033) & 845 & 19 & 0.022 & & & \\
%& & & &    \\
\multicolumn{8}{l}{\bf with prior + with extinction corr.}\\
%& & & &    \\
all     & 0.033 (0.025) &  868& 19  &   0.022 & 0.20 & 0.63 & 0.17 \\
$<0.2$  & 0.031 (0.025) & 838 & 15 & 0.018 & & & \\
$<0.3$  & 0.031 (0.025) & 846 & 16 & 0.019 & & & \\
%& & & &    \\
\multicolumn{8}{l}{\bf with prior + with extinction corr.}\\
%& & & &    \\
%& & & &    \\
Ellipticals & 0.034 (0.028) & 174 & 5 & 0.029  & & & \\
Spirals     & 0.030 (0.023) & 543 & 10 & 0.018 & & & \\
Starbursts  & 0.042 (0.027) & 151 & 4 & 0.026 & & &\\
AGNs        & 0.10 (0.026) & 12  & 1 & 0.083 & & & \\

\tablenotetext{1}{{\it rms} $\sigma(\Delta(z))$ values for all the spectroscopic sample and for the samples defines based on $D_{95}$ parameters. The values in brackets are the {\it rms} values measured with the outliers removed.}
\tablenotetext{2}{Total number of objects with $D_{95} < D_{95}^0$}
\tablenotetext{3}{Number of outliers in the $D_{95} < D_{95}^0$ sample. This is defined as $\Delta(z) > 3 \sigma(\Delta(z))$}

\end{tabular} 
\end{table*}

\clearpage
\begin{table*}
\caption[]{List of the initial parameters used for  different codes }

\begin{tabular}{ l|ccc||ccc|}

 &\multicolumn{3}{c}{without priors} & \multicolumn{3}{c}{with priors} \\
&ML  & SED optimization& Reddening & Baysian  & SED optimization& Reddening\\
\hline
BPZ           & X                       & X\tablenotemark{1} & -- &X& X\tablenotemark{1} & --\\
COSMOS & Best $\chi^2$ & X\tablenotemark{1} & X & X&X\tablenotemark{1}   & X \\
Le Phare  & X                       & X         & X & X&X             & X \\
ZEBRA     & X                       & X         & --  & X& X            & -- \\
\end{tabular} \label{setup codes}

\tablenotetext{1}{The optimization of the SED has been done externally to the codes.}
\end{table*}

\begin{table}
\caption{Accuracy of the codes with {\it  no priors } compared to the
spectroscopic sample and compared with the others}
\begin{center}
\begin{tabular}{c||cc|cc|cc|cc|}
no priors&\multicolumn{2}{|c|}{COSMOS}&\multicolumn{2}{|c|}{Le
Phare}&\multicolumn{2}{|c|}{BPZ}&\multicolumn{2}{|c|}{ZEBRA}\\
\hline
\hline
&-0.030 & 0.028 &-0.024&0.032&-0.030&0.027&-0.022&0.024\\
Z$_{spec}$&\multicolumn{2}{|c|}{\bf 0.029}&\multicolumn{2}{|c|}{\bf 0.028}&\multicolumn{2}{|c|}{\bf 0.029}&\multicolumn{2}{|c|}{\bf 0.023}\\
\hline
&&  & -0.026&0.036&-0.026&0.026&-0.027&0.024\\
COSMOS&&&\multicolumn{2}{|c|}{0.031}&\multicolumn{2}{|c|}{0.026}&\multicolumn{2}
{|c|}{0.026}\\
\hline
&& & & & -0.026&0.019&-0.027&0.023\\
Le Phare&&&&&\multicolumn{2}{|c|}{0.022}&\multicolumn{2}{|c|}{0.025}\\
\hline
&&& & & &&-0.022&0.020\\
BPZ&&&&&&&\multicolumn{2}{|c|}{0.021}\\
\hline
\hline
\end{tabular}
\end{center}
\label{sigma codes_01}
\end{table}

\begin{table}
\caption{Accuracy of the codes whit {\it priors } compared to the
spectroscopic sample and compared with the others}
\begin{center}
\begin{tabular}{c||cc|cc|cc|cc|}
priors&\multicolumn{2}{|c|}{COSMOS}&\multicolumn{2}{|c|}{Le
Phare}&\multicolumn{2}{|c|}{BPZ}&\multicolumn{2}{|c|}{ZEBRA}\\
\hline
\hline
&-0.025 & 0.024 &-0.025&0.031&-0.030&0.026&-0.020&0.026\\
Z$_{spec}$&\multicolumn{2}{|c|}{\bf{0.025}}&\multicolumn{2}{|c|}{\bf{0.028}}&\multicolumn{2}{|c|}{\bf{0.028}}&\multicolumn{2}{|c|}{\bf{0.023}}\\
\hline
&&  & -0.030&0.022&-0.020&0.021&-0.024&0.014\\
COSMOS&&&\multicolumn{2}{|c|}{0.026}&\multicolumn{2}{|c|}{0.021}&\multicolumn{2}
{|c|}{0.019}\\
\hline
&& & & & -0.017&0.024&-0.025&0.024\\
Le Phare&&&&&\multicolumn{2}{|c|}{0.020}&\multicolumn{2}{|c|}{0.024}\\
\hline
&&& & & &&-0.025&0.016\\
BPZ&&&&&&&\multicolumn{2}{|c|}{0.020}\\
\hline
\hline
\end{tabular}
\end{center}
\label{sigma codes_02}
\end{table}

\clearpage

\begin{figure}
%\epsscale{0.8}
\includegraphics[angle=90,scale=0.8]{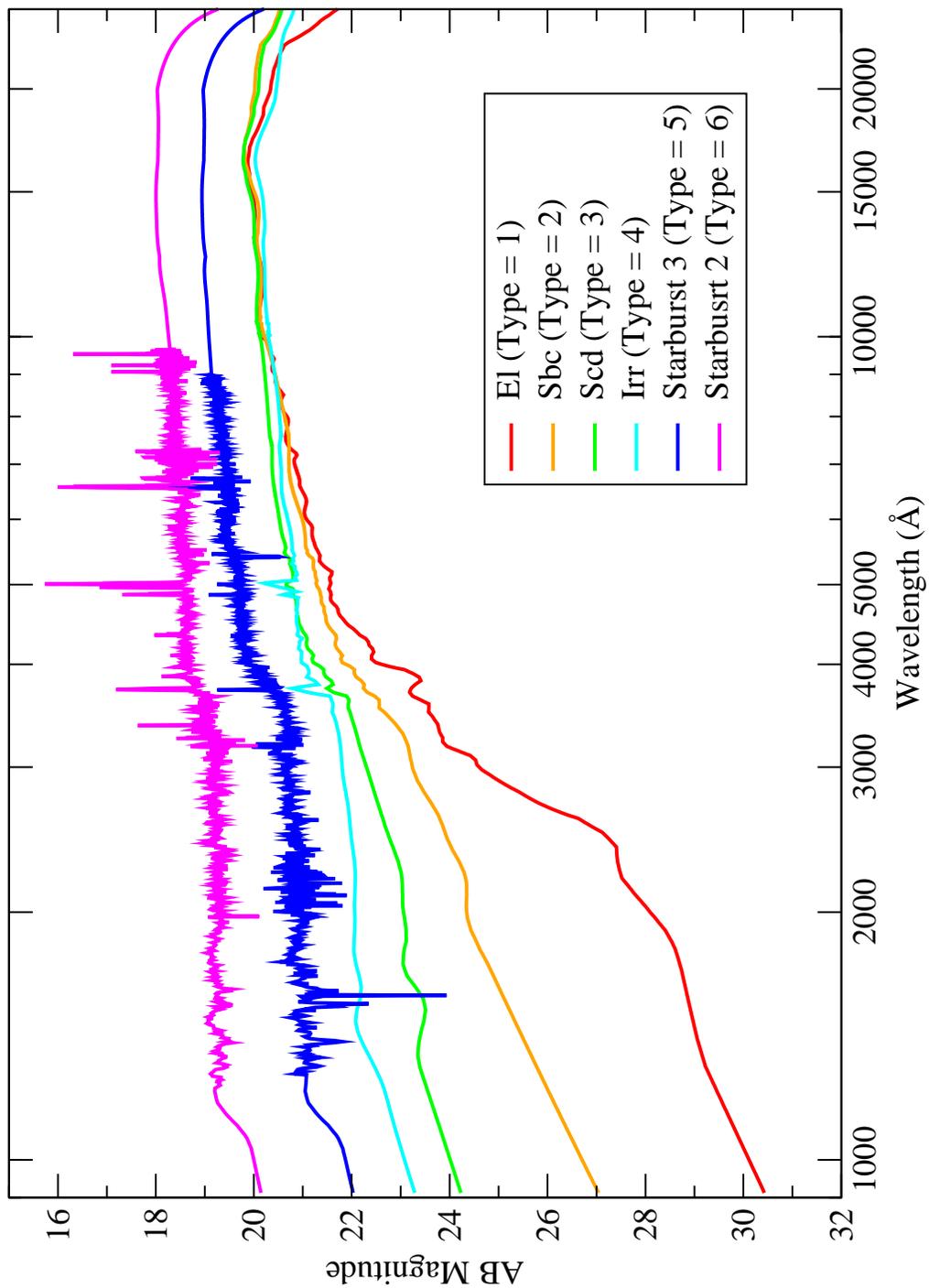}
\caption{Spectral Energy Distributions used as templates for photometric 
redshift 
measurement. These are trained to minimise the residuals between the 
photometric and spectroscopic redshifts for a sample of galaxies in HDF-N 
with available such data.}
\end{figure}

\begin{figure}
%\epsscale{0.8}
\includegraphics[angle=0,scale=1]{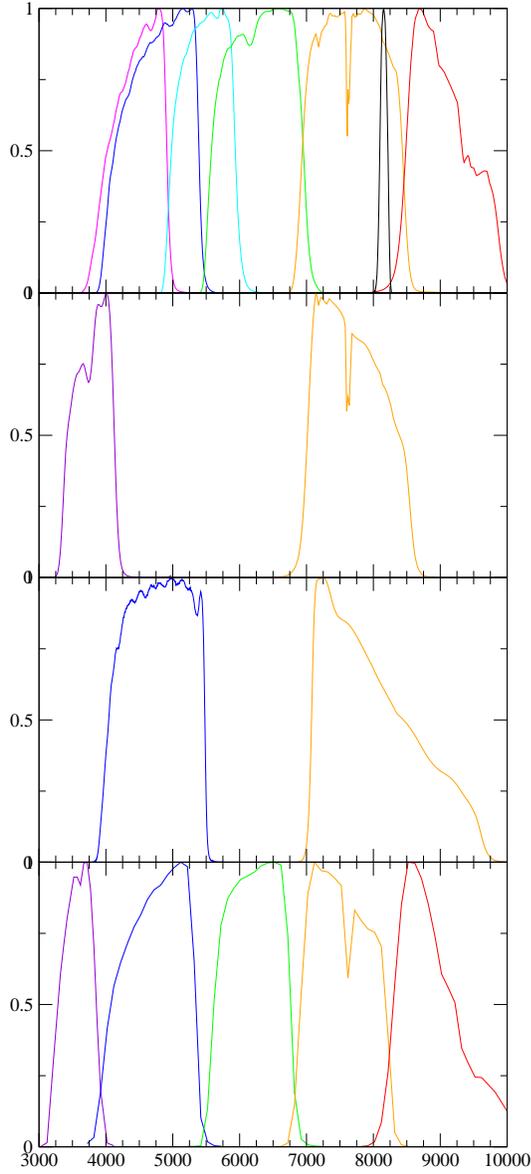}
\caption{Total Response Functions for the filters used in photometric 
observations and photometric redshift measurement of the COSMOS. The filters
consist of: top panel: B$_j$V$_jg^+r^+i^+z^+$ and $NB816$ (Subaru/SupremeCam);
second panel: u$^*$ and $i^*$ (CFHT); Third panel: B$_j$ and I$_c$; Forth panel: u'g'r'i'z' (SDSS) }
\end{figure}

\begin{figure}
%\epsscale{0.8}
\includegraphics[angle=0,scale=0.8]{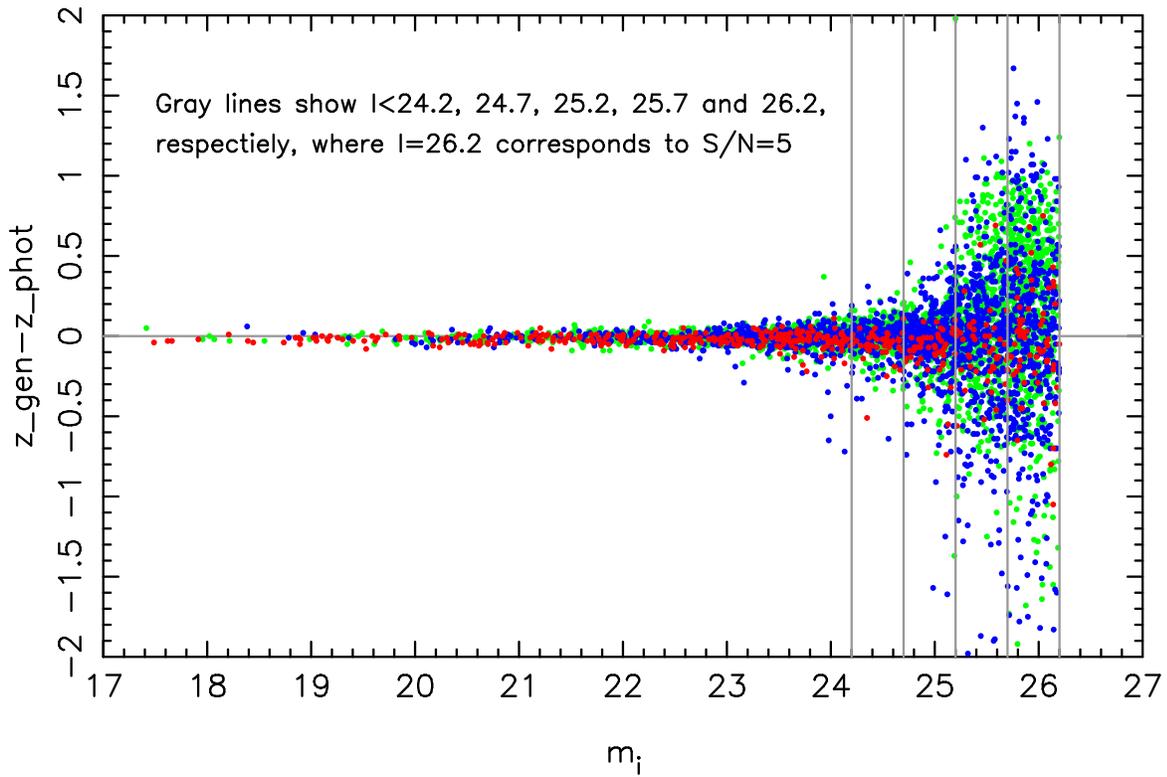}
%\plotone{junk3.ps}
\caption{Simulation presenting the comparison between the input and output 
redshifts as a function of magnitude limit, S/N ratios and spectral types 
(elliptical (red), early/intermediate type spirals (green); late-type and
starburst galaxies (blue).}
\end{figure}

\begin{figure}
%\epsscale{0.8}
\epsscale{1.0}
\includegraphics[angle=0,scale=0.8]{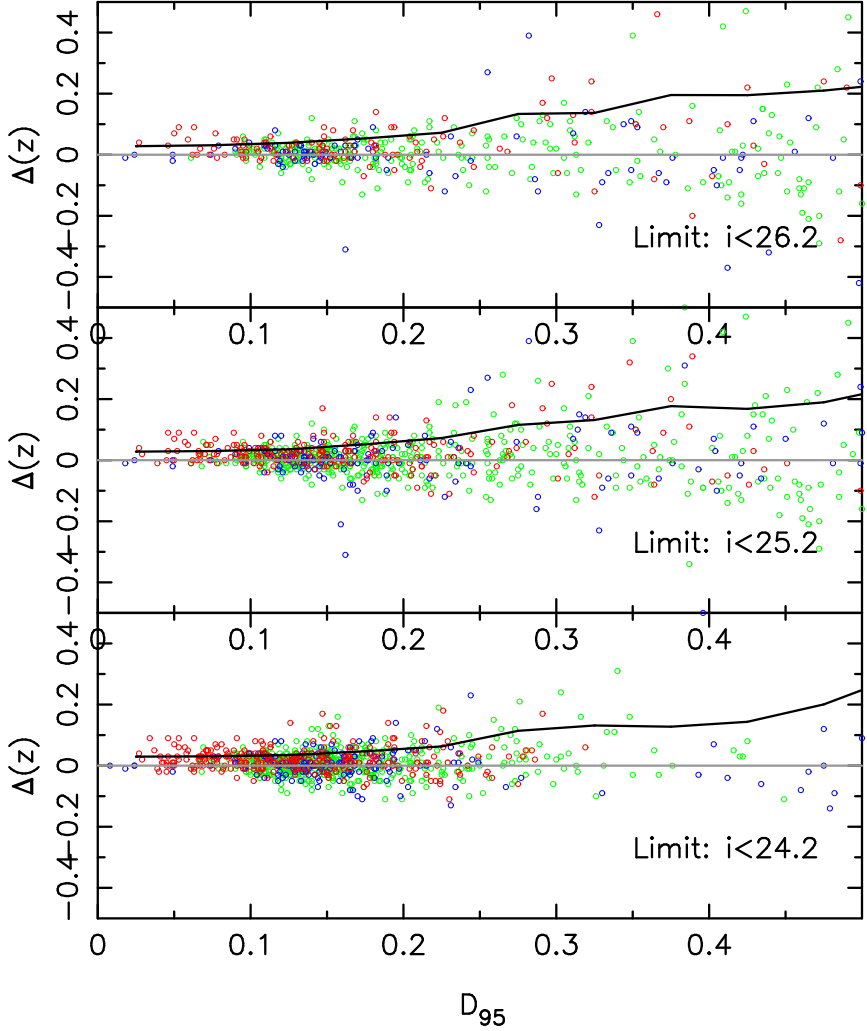}
\caption{Simulation results for different magnitude limits 
demonstrating dependence of $D_{95}$ on  $\Delta(z)$, which is a measure
of the accuracy of the estimated photometric redshifts.  
The scatter in $\Delta(z)$ increases towards
higher $D_{95}$ values and fainter magnitude limits. The black line
shows variation in {\it rms} for $\Delta(z)$ as a function of $D_{95}$. 
For clarity, we only present the plots for $0 < D_{95} < 0.5$. A number 
of points on the $i < 26.2$ panel scatter beyond the above $D_{95}$ and
$\Delta(z)$ range, as they are undetected in the short wavelength bands and photometric redshift get less reliable . This is the reason for a relatively smaller number of 
points on the $i < 26.2 $ mag. panel.}
\clearpage
%\includegraphics[angle=0,scale=0.8]{deltaz_d95_2.ps}
%\plotone{junk3.ps}

\end{figure}

\begin{figure}
%\epsscale{0.8}
\includegraphics[angle=0,scale=0.8]{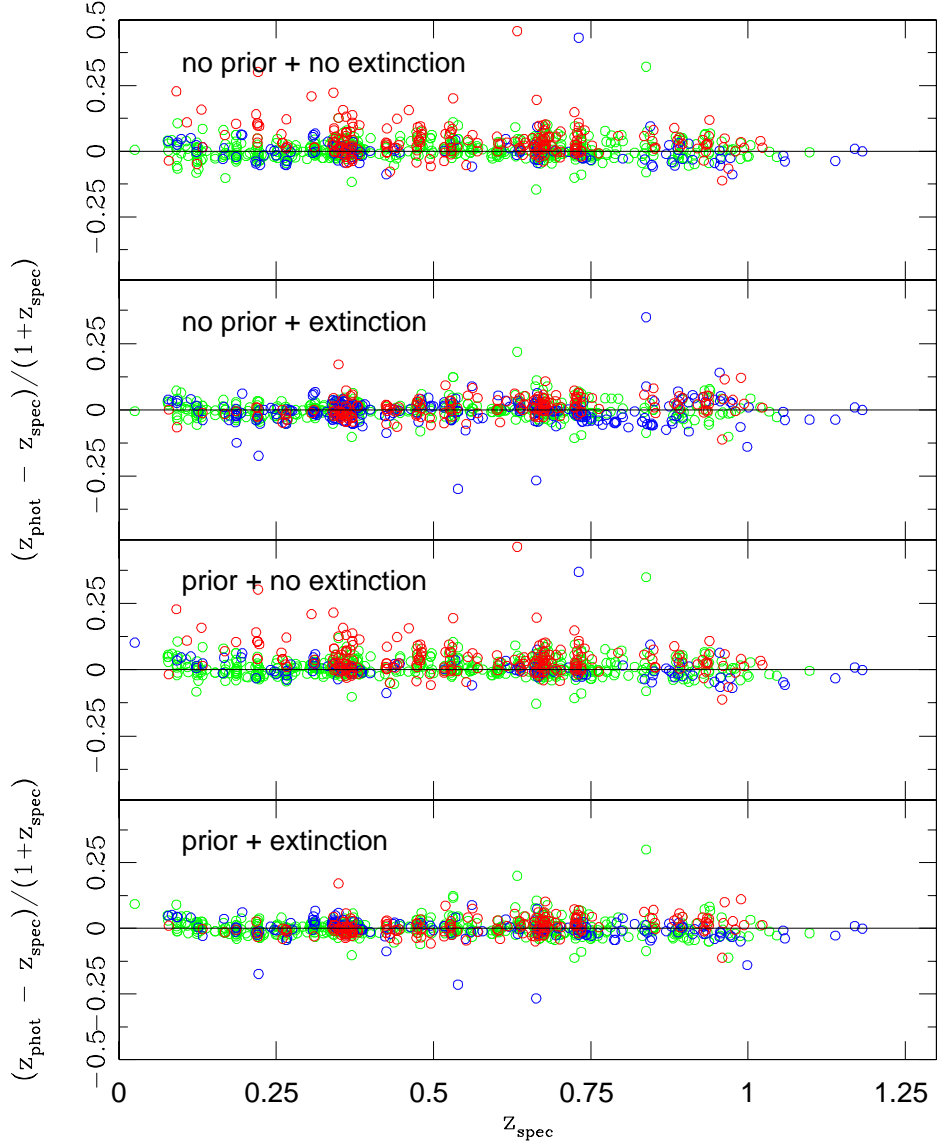}
\caption{Comparison between photometric and spectroscopic redshifts for 
a sample of 958 galaxies in COSMOS with available spectroscopic redshifts.  
The colors correspond to elliptical (red), spiral (green) and starburst (blue) 
spectral types. The spectral types are evenly distributed with redshift, 
indicating no bias in spectral type classification as a function of redshift.
The smallest scatter in $\Delta(z)$ ($(z_{phot}-z_{spec})/(1+z_{spec})$) obtained for the case including the prior
and extinction.}

\end{figure}

\begin{figure}
%\epsscale{0.8}
\includegraphics[angle=0,scale=0.8]{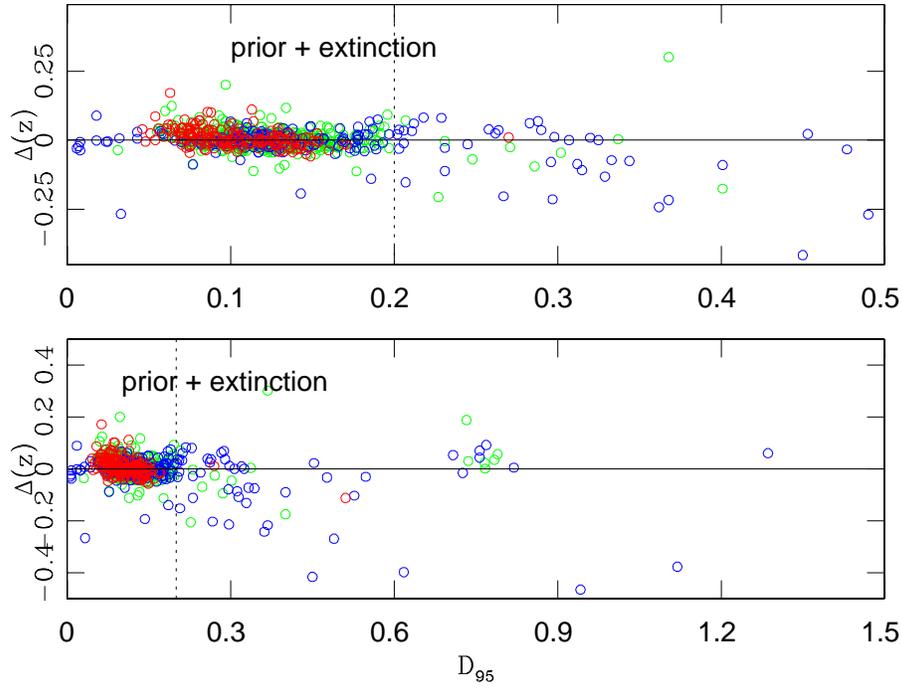}
\caption{{\bf (a).} Changes in $D_{95}$ parameter as a function of 
$\Delta(z)=(z_{phot}-z_{spec})/(1+z_{spec})$ for the spectroscopic
sample. Galaxies with $D_{95} < 0.2$ 
have more accurate photometric redshifts, as shown by the dotted line-  
Ellipticals (red), Spirals (green), starbursts (blue). 
{\bf (b).} Distribution of $D_{95}$ parameter for the spectroscopic sample. 
Different spectral types are identified with the color code as above.}
\end{figure}
\clearpage
\centerline{\includegraphics[angle=0,scale=0.8]{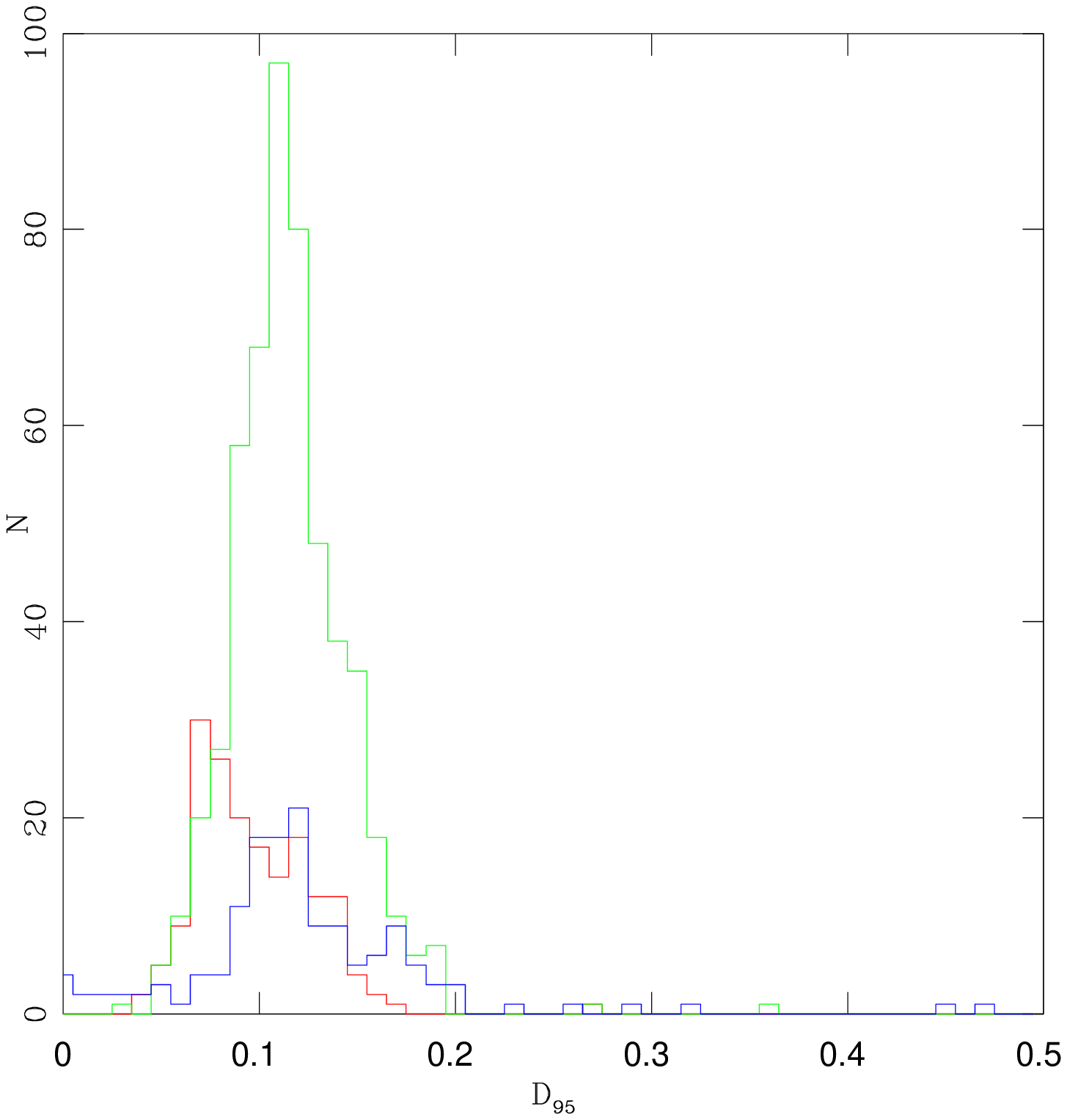}}
\clearpage

\begin{figure}
%\epsscale{0.8}
\includegraphics[angle=0,scale=0.8]{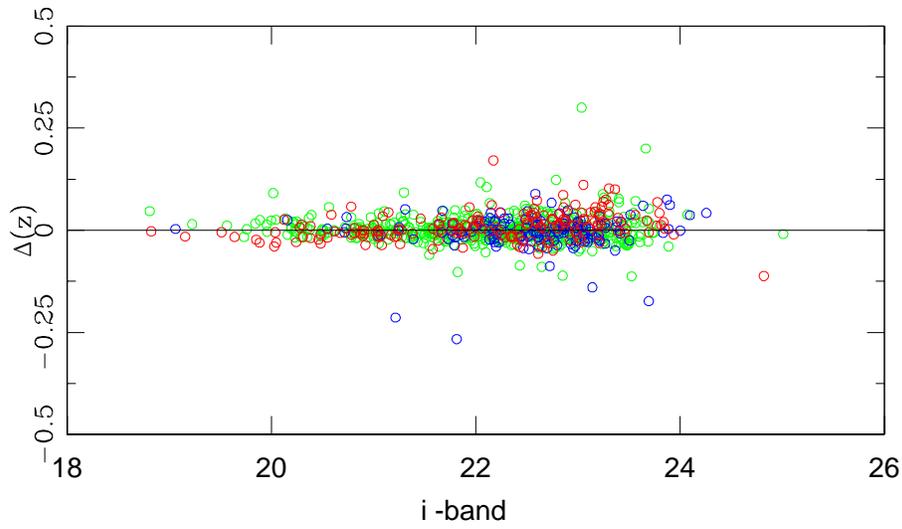}
\caption{Changes in  $\Delta(z)=(z_{spec}-z_{phot})/(1+z_{spec})$ as a 
function of $i-$ band magnitudes. There is slight increase in the scatter in $\Delta(z)$
(more uncertain photometric redshifts) at $i > 23$.}
\end{figure}

\begin{figure}
%\epsscale{0.8}
\includegraphics[angle=0,scale=0.8]{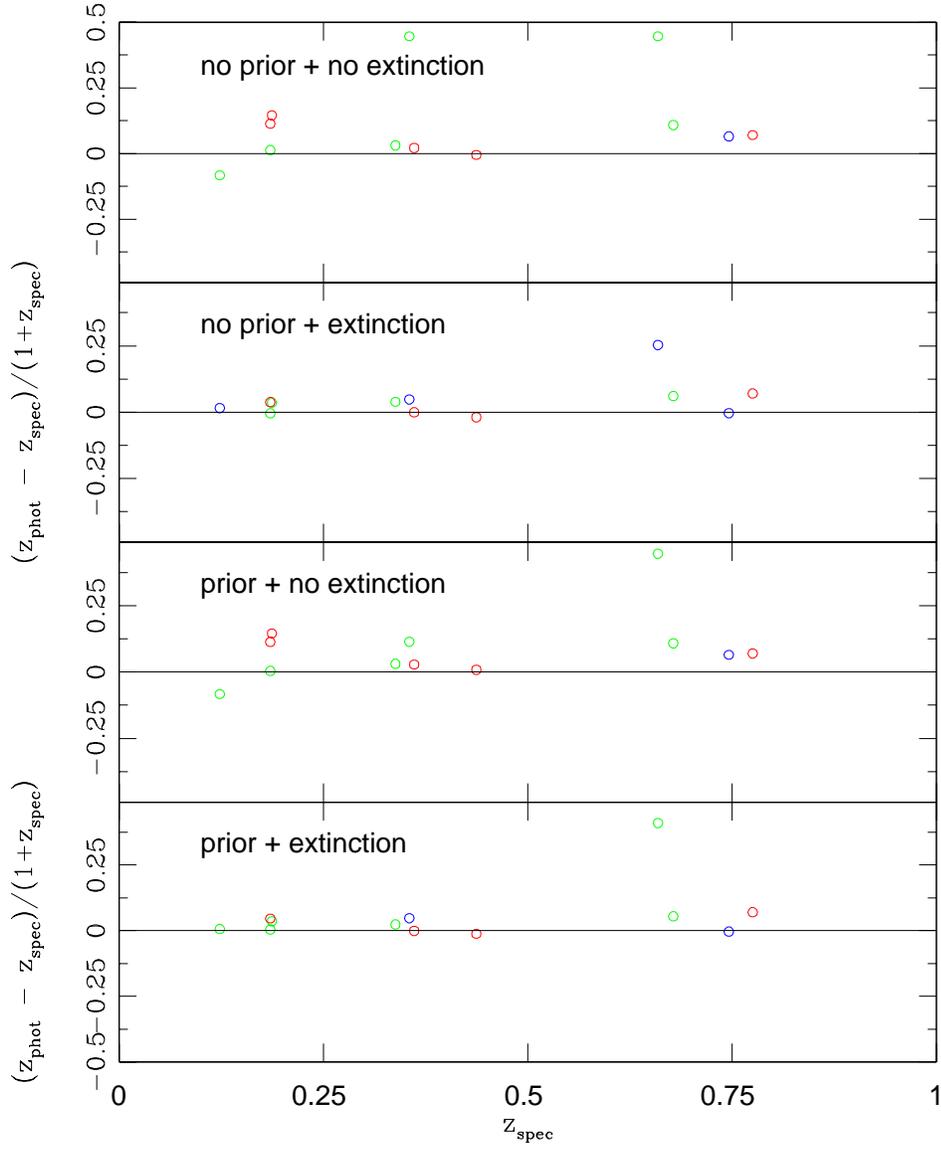}
\caption{The same as in Figure 5 but for 12 spectroscopically identified AGNs.}
\end{figure}

\begin{figure}
%\epsscale{0.8}
\includegraphics[angle=0,scale=0.8]{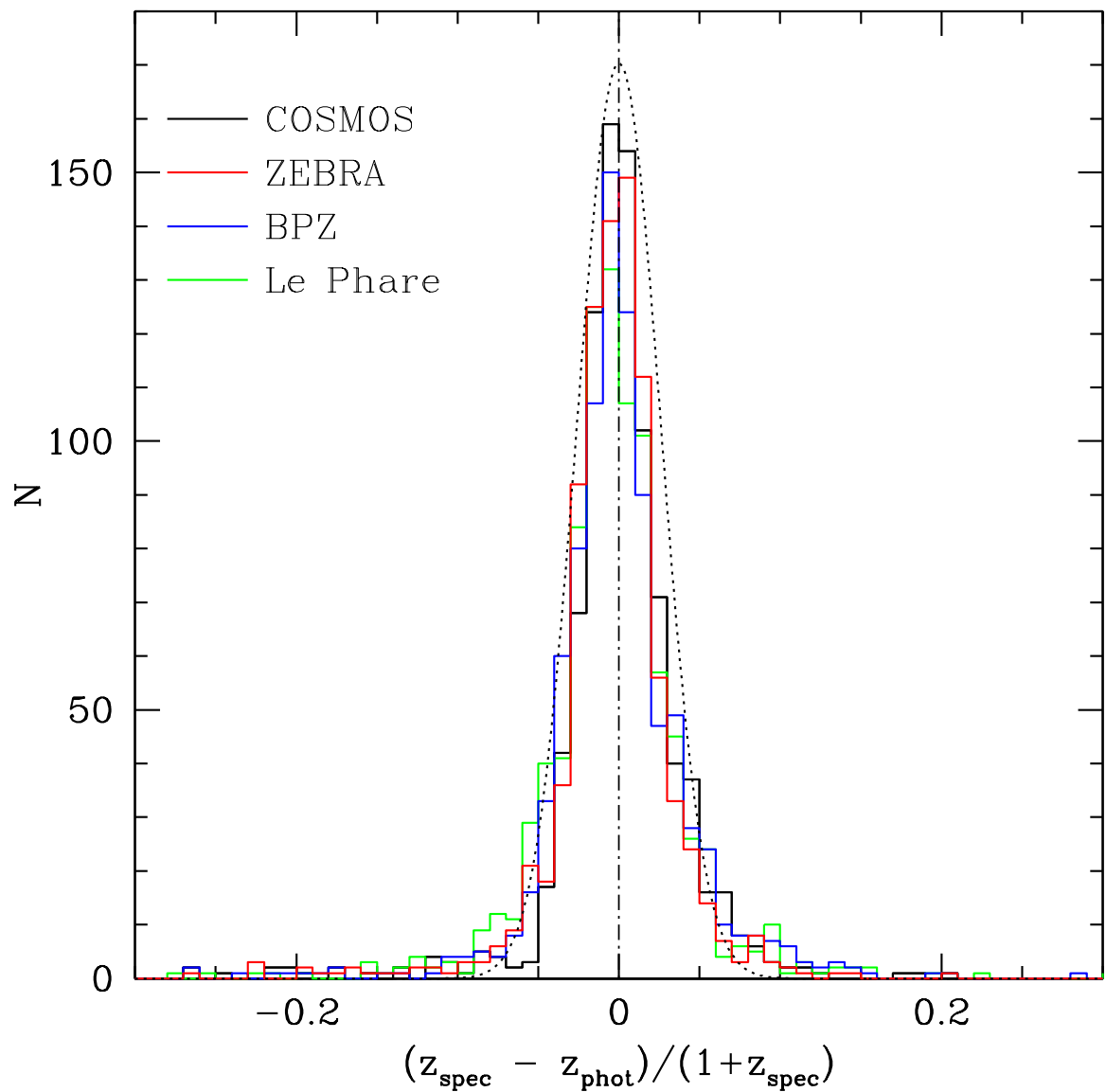}
\caption{Distribution of $\Delta(z)=(z_{spec}-z_{phot})/(1+z_{spec})$ values
from COSMOS, Le Pahre, ZEBRA and BPZ photometric redshift codes. All follow
a Gaussian distribution with a peak at $\Delta(z) \sim 0$. The distributions
are best fit with a Gaussian with $\sigma=0.026$. Photometric redshifts are
estimated assuming no priors.}
\end{figure}

\begin{figure}
%\epsscale{0.8}
\includegraphics[angle=0,scale=0.8]{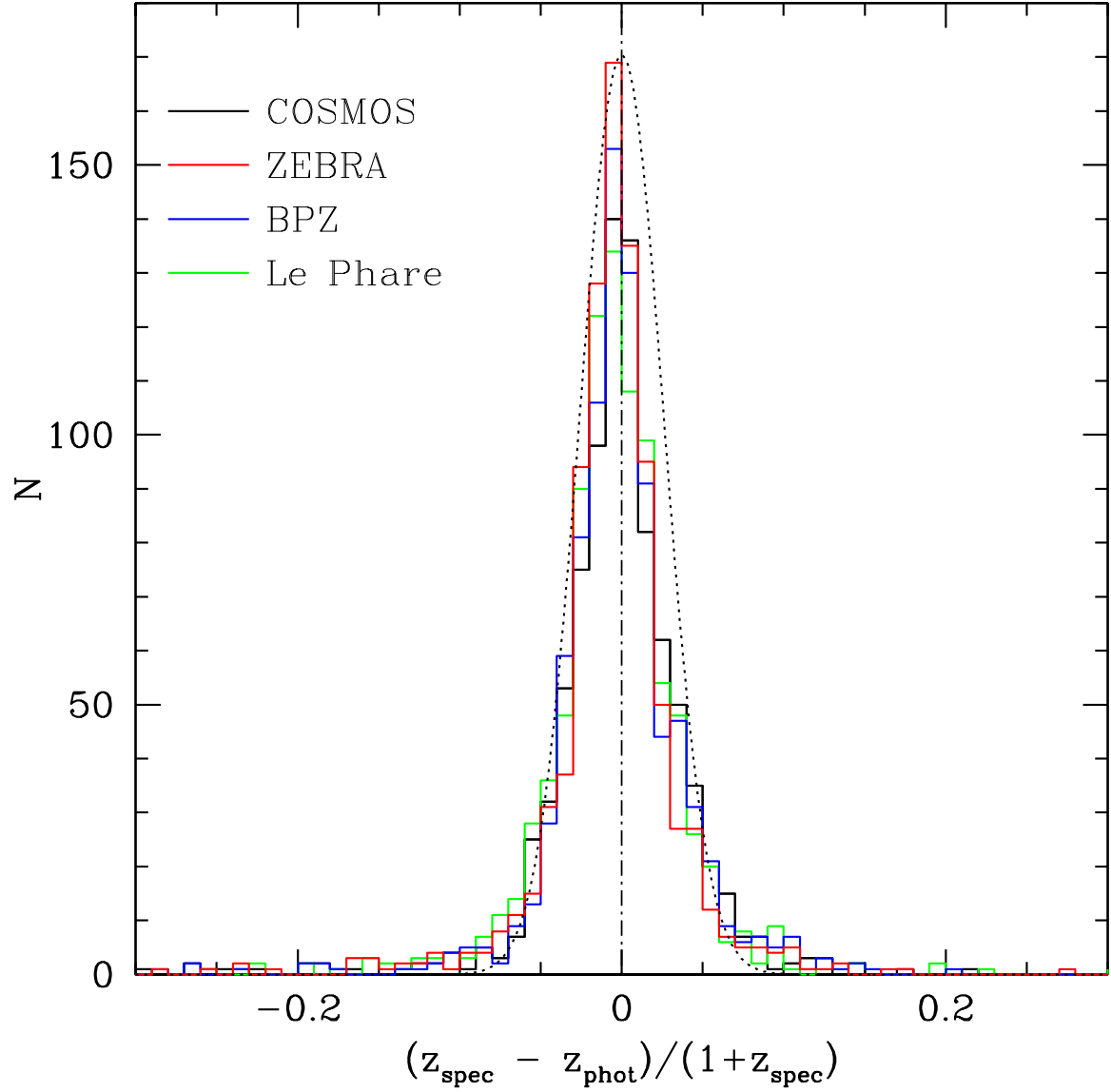}
\caption{The same as Figure 9 but assuming priors in estimating photometric redshifts.}
\end{figure}

\begin{figure}
%\epsscale{0.8}
\includegraphics[angle=0,scale=0.8]{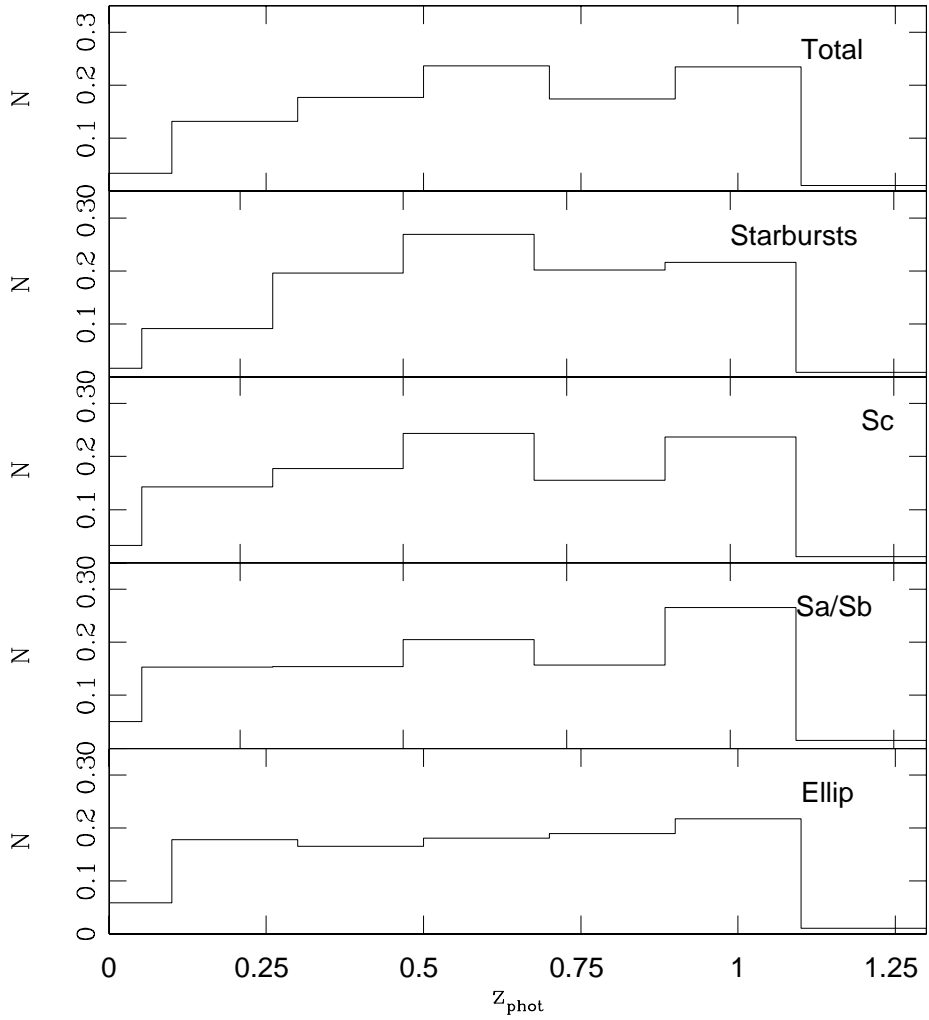}
%\plotone{phot_spec.eps}
\caption{Photometric Redshift distributions for different spectral types of galaxies for the entire COSMOS galaxies with $ i < 25$. The distributions for each spectral type are normalized to the total number of galaxies with that spectral type.}
\end{figure}

\begin{figure}
%\epsscale{0.8}
\includegraphics[angle=0,scale=0.8]{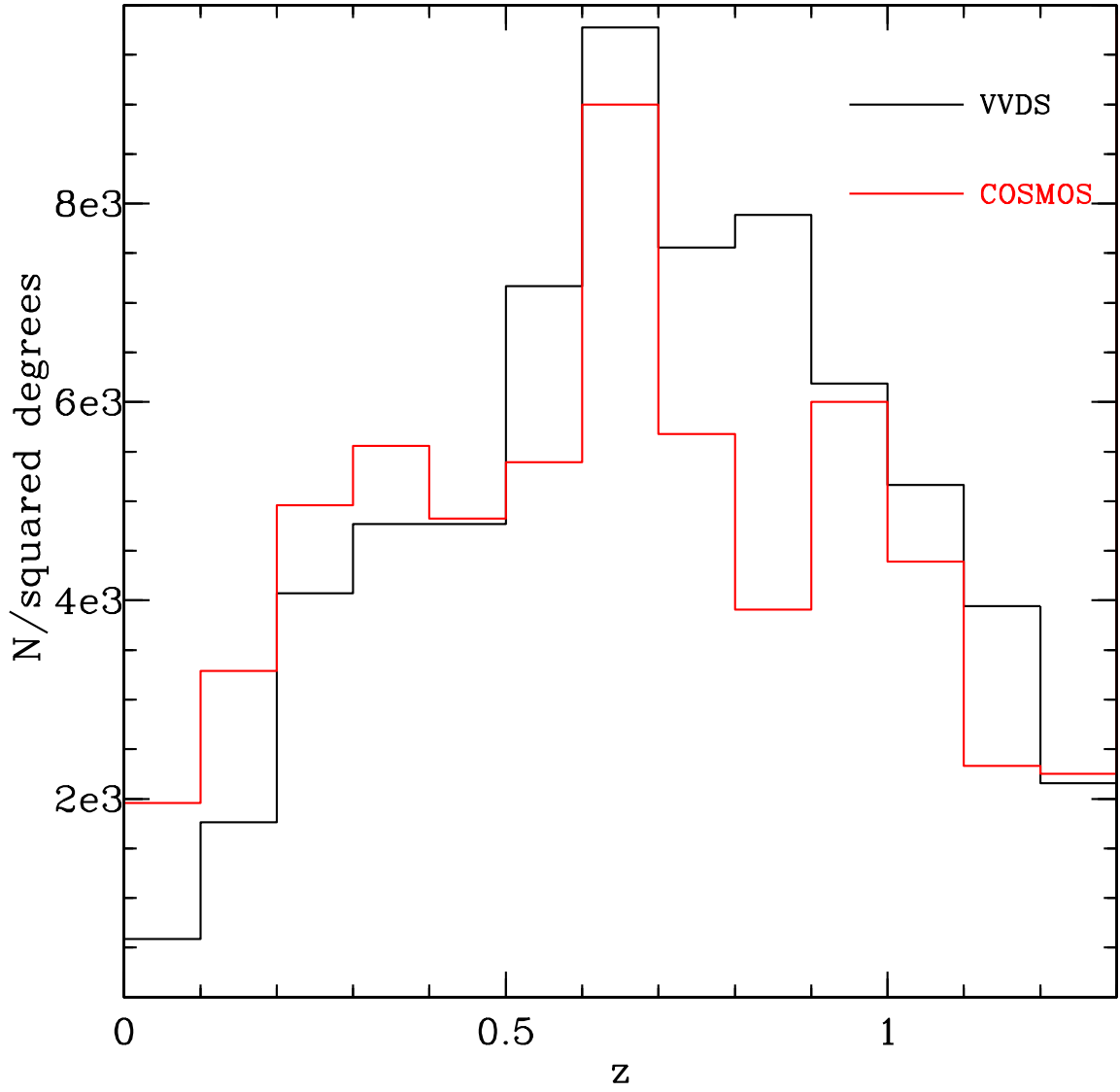}
\caption{Comparison between photometric redshift distribution from COSMOS
and spectroscopic redshift distribution (from VVDS). Galaxies to $i_{AB}\sim 24$ are used.}
\end{figure}

\begin{figure}
%\epsscale{0.8}
\includegraphics[angle=0,scale=0.8]{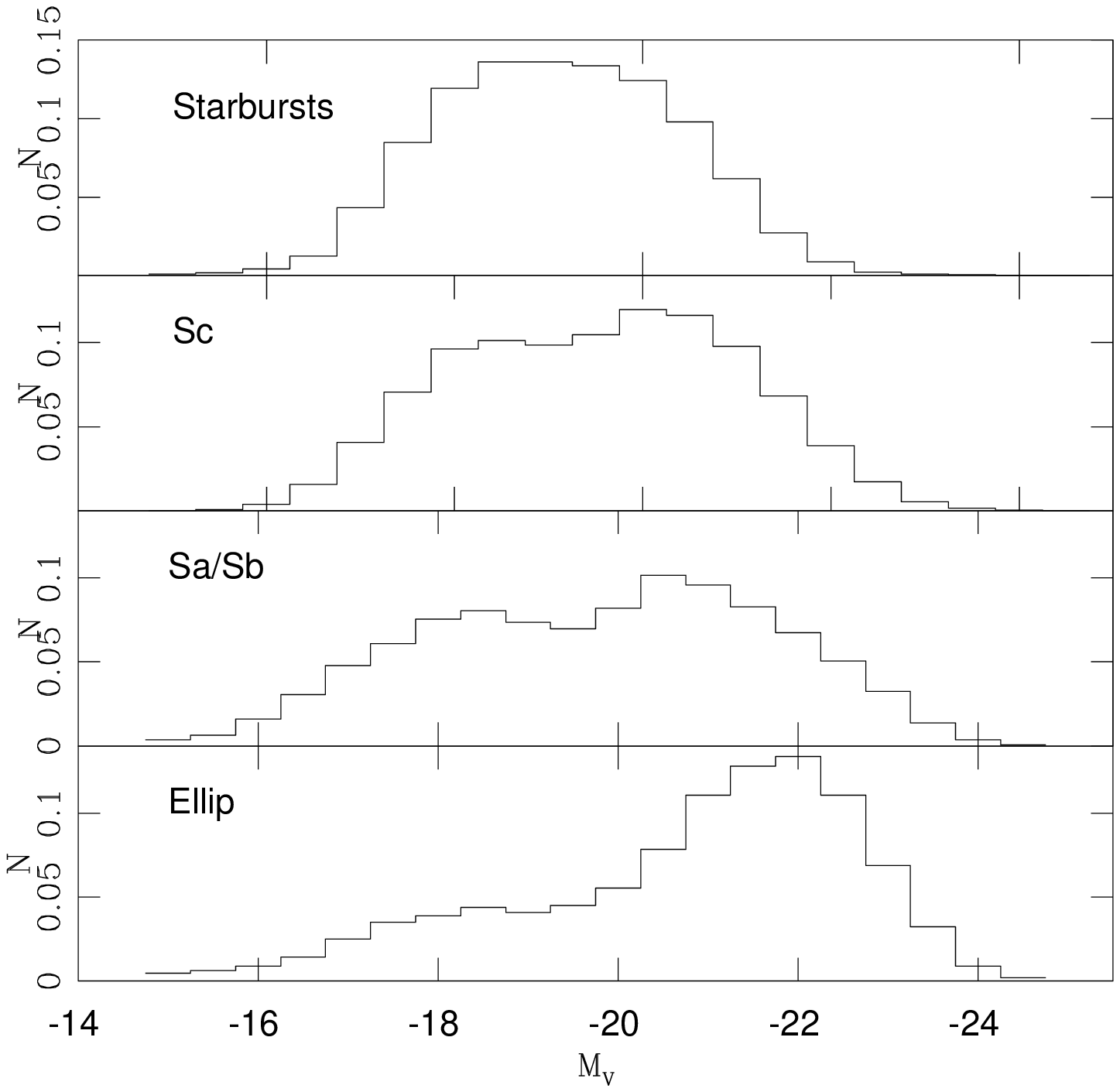}
%\plotone{phot_spec.eps}
\caption{ Rest-frame absolute magnitude distributions for different spectral types of galaxies in the entire COSMOS catalog. The expected trend is present, with early-type galaxies having brighter absolute magnitudes. The distributions for each spectral type are normalized to the total number of galaxies with that spectral type.}
\end{figure}

\begin{figure}
%\epsscale{0.8}
\includegraphics[angle=0,scale=0.8]{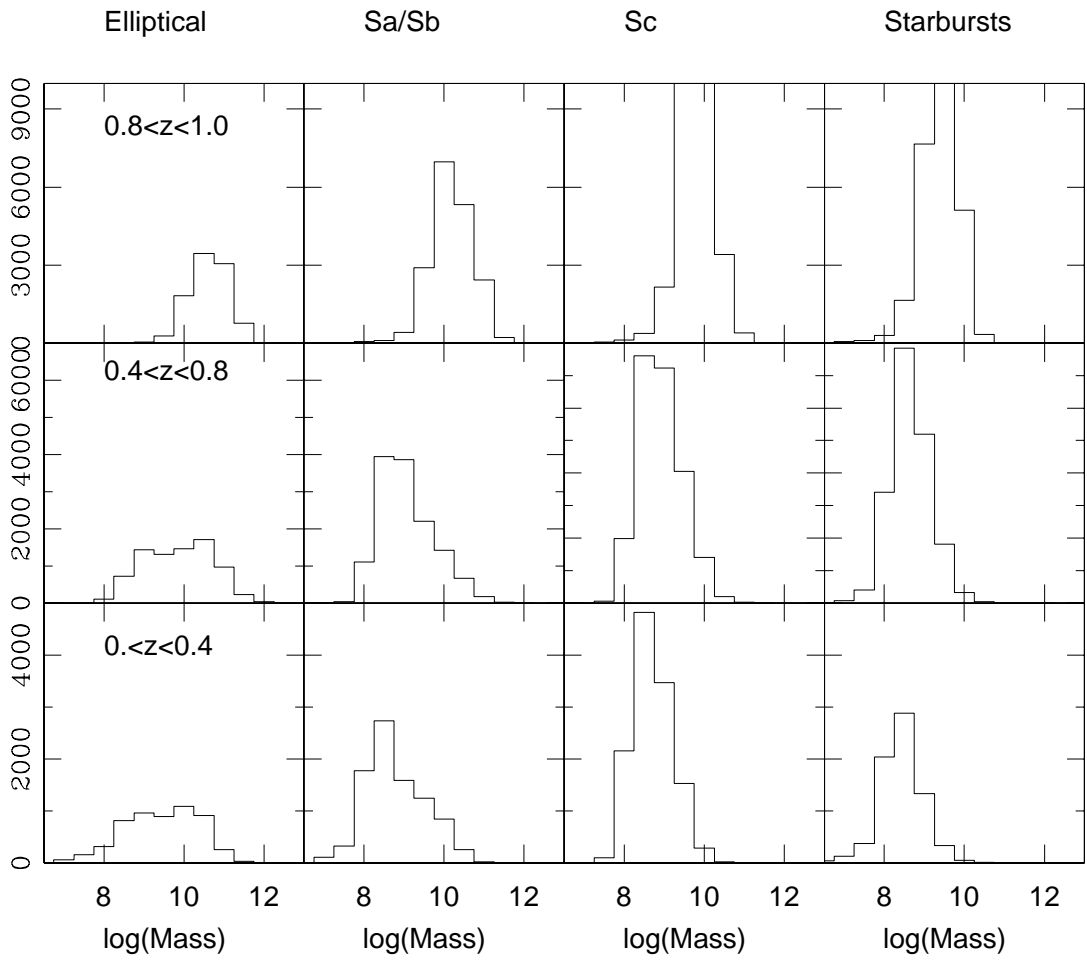}
%\plotone{phot_spec.eps}
\caption{Distribution of stellar mass as a function of spectral type and redshift for galaxies in the COSMOS survey}
\end{figure}

\end{document}